\documentclass[a4paper,10pt,openany]{article}
\usepackage{setspace}
\usepackage{hyperref}
\usepackage{pst-3d}
\usepackage{pst-3dplot}
\usepackage[utf8]{inputenc}
\usepackage[english]{babel}
\usepackage{amsmath,amssymb,amsthm,mathrsfs,amsfonts,dsfont}
\usepackage{graphicx}
\usepackage[small, skip=10pt]{caption}
\usepackage{epigraph}
\usepackage{indentfirst}
\usepackage{booktabs}
\usepackage{fancyhdr}
\usepackage{vmargin}
\usepackage{feynmf}
\usepackage{yfonts}
\usepackage{lscape}
\usepackage{textcomp}
\usepackage{pstricks}
\usepackage[metapost]{mfpic}
\usepackage{fancybox}
\usepackage{slashed}
\usepackage[verbose]{wrapfig}
\usepackage{pifont}
\usepackage{keystroke}
\usepackage{appendix}
\usepackage{enumitem}
\usepackage{array}
\usepackage{colortbl}
\usepackage{alltt}
\usepackage{boxedminipage}
\usepackage{tikz}
\usepackage{calc}
\usepackage{subfloat}
\usepackage{multirow}
\usepackage{cmll}
\usepackage{multicol}
\definecolor{color1}{RGB}{204,0,51}
\definecolor{color2}{RGB}{159,182,205}
\usepackage{bookmark}

\usepackage{verbatim}
\usepackage[vcentermath]{youngtab}
\usepackage{young}
\usepackage{transparent}
\usepackage{color,soul}
\usepackage{cancel}
\usepackage{bm}
\usepackage{pifont}
\usepackage{placeins}
\usepackage{mathtools}
\usepackage{physics}
\usepackage{rotating}
\usepackage[refpage]{nomencl}
\usepackage{hhline}
\usepackage{marginnote}
\usepackage{upgreek}
\usepackage{stackengine}
\usepackage{ytableau}
\usepackage{listings}
\usepackage{calligra}
\usepackage{afterpage}% for "\afterpage"
\usepackage{cite}
\usepackage{cleveref}
\usepackage{makecell}
\usepackage{accents}
\usepackage{breqn}
\usepackage{environ}
\usepackage{tikz-cd}
\tikzset{
    marrow/.style={decoration={markings,mark=at position 0.5 with {\arrow{#1}}}, postaction=decorate}
}

\numberwithin{equation}{section}

\usepackage{comment}

\usepackage{caption}
\usepackage{subcaption}
\usetikzlibrary{arrows}

\usepackage{xcolor}
\usetikzlibrary{calc}

\definecolor{darkergreen}{rgb}{0.0, 0.5, 0.0}

\allowdisplaybreaks

\sethlcolor{yellow}
\definecolor{boh}{RGB}{79,47,79}

\hypersetup{
        linkbordercolor={red}
}

\hypersetup{
unicode=false,          % non-Latin characters in Acrobat’s      bookmarks
pdftoolbar=true,        % show Acrobat’s toolbar?
pdfmenubar=true,        % show Acrobat’s menu?
pdffitwindow=false,     % window fit to page when opened
pdfstartview={FitH},    % fits the width of the page to the window
pdftitle={Generalized Newton-Cartan Geometries for Particles and Strings},    % title
pdfauthor={Eric Bergshoeff, Kevin van Helden, Johannes Lahnsteiner, Luca Romano, Jan Rosseel},     % author
pdfsubject={},   % subject of the document
pdfnewwindow=true,      % links in new window
colorlinks=false,       % false: boxed links; true: colored links
linkcolor=red,          % color of internal links
citecolor=red,        % color of links to bibliography
filecolor=red,      % color of file links
urlcolor=red           % color of external links
linkbordercolor={red},
citebordercolor={red},
urlbordercolor={red}
}

\makeatletter

\newcommand{\Rmnum}[1]{\expandafter\@slowromancap\romannumeral #1@}
\makeatother

\usepackage{alphalph}

\makeatletter
\newalphalph{\aalphalph}[mult]{\alphalph@alph}{26}
\newcommand{\alphalphval}[1]{%
\@ifundefined{c@#1}{% check first if #1 is a counter (\c@#1)
\aalphalph{#1}% No, it's most likely the direct value
}{%
\aalphalph{\value{#1}}% It's a counter, so use \value{#1}
}
}
\makeatother

\usepackage{color}

\def\chapterautorefname~#1\null{Chap.~(#1)\null}
\def\sectionautorefname~#1\null{Sec.~(#1)\null}
\def\subsectionautorefname~#1\null{sub--Sec.~(#1)\null}
\def\figureautorefname~#1\null{Fig.~(#1)\null}
\def\tableautorefname~#1\null{Tab.~(#1)\null}
\def\equationautorefname~#1\null{eq.~(#1)\null}

\def\equationautorefname~#1\null{eq.~(#1)\null}

\NewEnviron{multieq}[1][2]{

\begin{multicols}{#1}
\begin{subequations}
\setlength{\abovedisplayskip}{-15pt}
\allowdisplaybreaks
\begin{align}
\BODY
\end{align}
\end{subequations}
\end{multicols}
\setlength{\parindent}{0pt}
}

\DeclareMathAlphabet\mathbfcal{OMS}{cmsy}{b}{n}

\usepackage[yyyymmdd]{datetime}

\title{A conformal approach to matter coupled Aristotelian gravity}
\date{}

\begin{document}

\begin{flushright}
\small
%IFT-UAM/CSIC-19-10\\
%\texttt{arXiv:yymm.nnnnn [hep-th]}\\
July 22, 2025\\
%\textsuperscript{th}
\normalsize
\end{flushright}
{\let\newpage\relax\maketitle}
\maketitle
\def\equationautorefname~#1\null{eq.~(#1)\null}
\def\tableautorefname~#1\null{tab.~(#1)\null}

{\Large E.A. Bergshoeff$^1$, G. Giorgi$^2$, J. Rosseel$^3$ and  P. Salgado-Rebolledo$^4$,}

\vspace{0.5truecm}

\begin{center}

{\large $^1$ {\it Van Swinderen Institute, University of Groningen,}

{\it Nijenborgh 3, 9747 AG Groningen, The Netherlands}

\vspace{0.3truecm}

$^2$ {\it Departamento de F\' isica, Universidad de Murcia, }

{\it Campus de Espinardo, 30100 Murcia, Spain}

\vspace {0.3truecm}

$^3$ {\it Division of Theoretical Physics, Rudjer Bo\v skovi\' c Institute, }

{\it Bijeni\v cka 54, 10000 Zagreb, Croatia }

\vspace {0.3truecm}

$^4$ {\it Asia Pacific Center for Theoretical Physics (APCTP), }

{\it Pohang, Gyeongbuk 37673, Korea}

}
\end{center}

\vspace{2truecm}

\centerline{ABSTRACT}
\vspace{0.2truecm}

 We show how to take the first step in the conformal program for constructing general 
matter couplings to Aristotelian gravity with arbitrary $p$-brane foliation. For this purpose we extend the $p$-brane Aristotelian algebra to the direct sum of two conformal algebras: one with Minkowski signature for the longitudinal directions and a second one with Euclidean signature for the transverse directions. For some cases, it is sufficient to work with a subalgebra of this conformal extension that, instead of two dilatations that are isotropic in either the longitudinal or transverse directions, contains a single dilatation that acts on the longitudinal and transverse directions in an an-isotropic way. Using this conformal extension we show how different electric and magnetic versions of Aristotelian gravity can be constructed that all have the distinguishing property that they are not invariant under any (Galilean or Carrollian) boost symmetry. We next consider several matter couplings both for quadratic-derivative models as well as for some higher-derivative models that have recently been considered in connection with studies on fractons.

\newpage

\vspace{0.8cm}

\tableofcontents
\section{Introduction}

There exist many physical models that do not exhibit boost symmetry. Following \cite{Penrose:1968ar}, the geometry underlying these models is often called Aristotelian geometry. Aristotelian field theories have been studied in the context of non-relativistic naturalness, see, e.g.~\cite{Horava:2016vkl,Yan:2017mse}. More recently, hydrodynamic models without boost symmetry have been investigated in \cite{Novak:2019wqg,deBoer:2020xlc,Armas:2020mpr,Marotta:2023ayw,Grosvenor:2024vcn}, while Aristotelian G-structures appear in a wide range of physical systems such as Bjorken flow and defects in conformal field theory \cite{Gubser:2010ze,Chattopadhyay:2018apf,Billo:2016cpy,Arav:2024exg}. Aristotelian geometries have also occurred in the study of the GMP modes in the Fractional Quantum Hall Effect \cite{Gromov:2017qeb} and in the description of fractons \cite{Chamon:2004lew,Haah:2011drr,Pretko:2016kxt,Pretko:2018jbi} in a curved spacetime \cite{Slagle:2018kqf,Pena-Benitez:2021ipo,Bidussi:2021nmp,Jain:2021ibh,Pena-Benitez:2023aat,Hartong:2024hvs}. A more formal treatment of Aristotelian geometries can be found in \cite{Figueroa-OFarrill:2018ilb,Figueroa-OFarrill:2020gpr}. Recently, a string-foliated version of Aristotelian geometry has occurred in a study of multi-Weyl semimetals \cite{Ghosh:2025eob}. Following
\cite{Bergshoeff:2023rkk}, this naturally leads to the notion of a $p$-brane Aristotelian geometry.

In view of all these connections, it is the purpose of this paper to give a unified treatment of $p$-brane Aristotelian geometry and, following the conformal compensating program, see, e.g.~\cite{Freedman:2012zz}\,\footnote{For a short recent introduction to this technique, see section 2 of \cite{Bergshoeff:2024ilz}.},  to make the first step in constructing matter-coupled Aristotelian gravity theories. When following this program, it is not obvious which conformal extension of the Aristotelian geometry one should use. One of the criteria such a conformal extension should satisfy is that one can use it to construct an Aristotelian gravity theory without matter. It turns out that the basic building blocks of such an Aristotelian gravity theory are either a curvature tensor for a spin-connection or an intrinsic torsion tensor. These two building blocks should be contained in the expressions of one of the dependent conformal gauge fields. In practice, curvature tensors for a spin-connection can be found in the solution of dependent special conformal gauge fields whereas intrinsic torsion tensors can be found in the dependent expression of a dilatation gauge field. However, this can only happen if the commutators of the conformal Aristotelian algebra allow to solve for the relevant conformal gauge field. In the relativistic case for instance, the fact that the special conformal gauge field can be solved in terms of a Riemann curvature tensor is immediately related to the fact that the commutator of a spacetime translation with a special conformal transformation gives a Lorentz transformation. Because of this, the conformal curvature tensor for the Lorentz spin-connection contains the product of an invertible Vielbein field with a special conformal gauge field which allows one to solve for the special conformal gauge field in terms of the Riemann tensor by setting some components of the conformal curvature tensor equal to zero.

It turns out that the same thing happens in the Aristotelian case, i.e.~one can solve for a special conformal gauge field in terms of an Aristotelian curvature tensor, if one extends the $p$-brane Aristotelian algebra in $D$ dimensions to the following direct sum of two conformal algebras:
\begin{align}
  \label{eq:galg1}
  \mathfrak{so}(2,p+1) \oplus \mathfrak{so}(1,D-p) \,.
\end{align}
Here, $\mathfrak{so}(2,p+1)$ is the conformal algebra in $(p+1)$ dimensions with Minkowski signature and $\mathfrak{so}(1,D-p)$ is the conformal algebra in $D-p-1$ dimensions with Euclidean signature. This conformal extension contains two independent isotropic dilatation generators $D_1$ and $D_2$. We will refer to the algebra \eqref{eq:galg1} as the {\sl isotropic} conformal Aristotelian algebra. If we restrict ourselves to Aristotelian gravity theories that only contain intrinsic torsion tensors as basic building blocks and no Aristotelian connection curvatures it suffices to work with a subalgebra of \eqref{eq:galg1} that will be specified later in this paper. This subalgebra contains only one an-isotropic dilatation generator $D$ with critical exponent $z$ given by
\begin{equation}
D \equiv zD_1 + D_2\,.
\end{equation}
We will call this subalgebra the {\sl an-isotropic} conformal Aristotelian algebra.\footnote{We call the algebra \eqref{eq:galg1} isotropic because the two dilatation generators $D_1$, resp. $D_2$, act isotropically on the longitudinal, resp. transverse, directions only. By contrast, in the an-isotropic algebra the dilatation $D$ acts an-isotropically on the combined set of longitudinal and transverse directions. }

Different realizations of the concept of {\it gauge-equivalent formulations} play an important role in our construction of matter-coupled Aristotelian gravity theories. It works as follows.
 Suppose $\mathcal{L}(A)$ is a Lagrangian that depends on a single field $A$. Defining
\begin{equation}\label{redef1}
A \equiv B + C
\end{equation}
and substituting this definition back into $\mathcal{L}(A)$ we obtain a Lagrangian $\mathcal{L}(B,C)$ that depends on two scalars $B$ and $C$ and that is furthermore invariant under the following gauge transformation with gauge parameter $\epsilon(x)$:
\begin{equation}
\delta B = \epsilon(x)\,,\hskip 2truecm \delta C = -\epsilon(x)\,.
\end{equation}
This gauge symmetry guarantees that the fields $B$ and $C$ occur in the Lagrangian $\mathcal{L}(B,C)$ only in the combination $B+C$. We say that the Lagrangians $\mathcal{L}(A)$ and $\mathcal{L}(B,C)$ are {\it gauge-equivalent} to each other. One can go back to the original formulation in terms of $\mathcal{L}(A)$, e.g., by imposing the gauge-fixing condition $C=0$
and identifying $A\equiv B$.

In this paper, we will consider such gauge-equivalent formulations both for a real scalar $A=r$, where the corresponding gauge symmetries are the gravitational dilatations, as well as for a real scalar $A=\theta$ with corresponding gauge symmetry given by the gauged dipole symmetry of fractons. These two real scalar fields $r$ and $\theta$ can be combined into a complex scalar $\Phi = re^{i\theta}$.

As an example, we consider the Einstein-Hilbert Lagrangian.  Extending the Poincar\'e algebra to a relativistic conformal algebra with an isotropic dilatation symmetry, we replace the Poincar\'e Vielbein field
$E_\mu{}^{\hat A}$ in the Einstein-Hilbert Lagrangian
by the product of a conformal compensating\,\footnote{The scalar $r$ is called a {\sl compensating} scalar because it compensates for the dilatations of the conformal Vielbein field.} scalar $r$ with dilatation weight $w$ and a conformal Vielbein field $\big(E_\mu{}^{\hat A}\big)^C$ with weight 1 as follows:
\begin{equation}\label{redef2}
E_\mu{}^{\hat A} = r^{-1/w}\, \big(E_\mu{}^{\hat A}\big)^C\,.
\end{equation}
In this way one obtains the Lagrangian for a dynamical scalar $r$ coupled to relativistic conformal gravity. The relation \eqref{redef2} is the analogue of eq.~\eqref{redef1}. The Einstein-Hilbert action of general relativity can be re-obtained by
fixing the dilatations setting the conformal scalar $r$ to some constant. To obtain non-trivial matter couplings one should replace the single scalar by a function of $N$ scalars such that after gauge-fixing one is left with $N-1$ scalars coupled to gravity. In this paper we will perform similar manipulations replacing general relativity by Aristotelian gravity and the relativistic conformal algebra by an appropriate conformal extension of the Aristotelian algebra.

The compensating scalar field $\theta$ plays a similar role as the compensating scalar $r$ but instead of gravitational dilatations the scalar $\theta$ compensates for gauge symmetries. It is instructive to first consider an Abelian U(1) symmetry. Our starting point is the real dynamical scalar field $\theta(x)$ with  action
\begin{equation}\label{starting}
\mathcal{L} = -\tfrac{1}{2}\big(\partial_\mu\theta\big)^2\,.
\end{equation}
This action is invariant under the transformation $\delta\theta = m\Lambda$ for constant $m$ and constant symmetry parameter $\Lambda$. We can gauge this symmetry by replacing the parameter $\Lambda$ by $\Lambda(x)$ and introducing a gauge field $A_\mu$ with transformation rule $\delta A_\mu = \partial_\mu \Lambda(x)$.
This leads to the following gauge-invariant action
\begin{equation}\label{starting2}
\mathcal{L} = -\tfrac{1}{2}\big(D_\mu\theta\big)^2
\end{equation}
with covariant derivative $D_\mu\theta$ given by $D_\mu\theta = \partial_\mu\theta -mA_\mu$.
We see that in this action the scalar field $\theta$ and the gauge field $A_\mu$ only occur in the gauge-invariant combination $A_\mu - m^{-1} \partial_\mu\theta$. This leads to the natural definition of the gauge-invariant vector
\begin{equation}
W_\mu \equiv A_\mu -\frac{1}{m}\partial_\mu\theta
\end{equation}
in terms of which the action can be rewritten as a mass term of the vector field $W_\mu$.
Adding to the gauge-invariant action \eqref{starting2} a kinetic term for the gauge field $A_\mu$
leads to a  Proca action for the massive vector field $W_\mu$.

A similar mechanism works for the   dipole symmetries that we consider in this paper except that we will gauge these dipole symmetries by introducing both a longitudinal vector gauge field $\phi_A$ with transformation rule $\delta \phi_A = \partial_A\Lambda(x)$ as well as a transverse symmetric tensor gauge field $A_{ab}$ with transformation rule
$\delta A_{ab} = \partial_a\partial_b \Lambda(x)$.\,\footnote{The reason that we use a symmetric tensor gauge field instead of a vector field for the transverse directions is that we wish to recover the algebra of dipole symmetries after truncating all gauge fields to zero. Such a gauging is possible if the model before gauging is invariant under both constant phase transformations and dipole symmetries.}
The compensating scalar $\theta$ and the gauge field $A_{ab}$ will then combine and lead to a mass term for the symmetric tensor \begin{equation}
W_{ab} = A_{ab} - \partial_a\partial_b\theta\,,
\end{equation}
which is the analogue of the relation \eqref{redef1}.

Following the above two examples we will consider in this work models with one or two dynamic scalars. Concerning the two-scalar models, we will consider both quadratic-derivative two-scalar models 
where the two scalars compensate for the two  dilatations of the isotropic conformal Aristotelian algebra as well as higher-derivative two-scalar models where one scalar 
compensates for the single dilatation of the an-isotropic conformal Aristotelian algebra and the second scalar compensates for the gauged dipole symmetries.

This paper is organized as follows. In section 2 we will discuss the $p$-brane Aristotelian geometries and their conformal extensions. In particular, we will discuss the Aristotelian intrinsic torsion tensors and show how they can be used to classify (conformal) Aristotelian geometries. Furthermore, we will discuss conformal Aristotelian gravity as the gauge theory of the conformal Aristotelian algebra. In section 3 we will show how examples of Aristotelian gravity theories without matter can be constructed using the conformal program. We will thereby distinguish between three classes of invariants: electric, magnetic and electric-magnetic. We will show how these different types of  Aristotelian gravity theories without matter can be constructed following the conformal program using {\sl two} compensating scalars corresponding to the two dilatations of the isotropic conformal Aristotelian algebra. Furthermore, we will show how the electric Aristotelian gravity theory can also be constructed following the conformal program based upon the an-isotropic conformal Aristotelian algebra using a single scalar that compensates for the an-isotropic dilatations. In section 4 we will use the an-isotropic conformal Aristotelian algebra to discuss matter coupled Aristotelian gravity theories involving scalar fields as well as vector/tensor gauge fields. We will do this both for quadratic-derivative models as well as for the higher-derivative models that have appeared in the context of fractons.
An outlook and possible extensions of our work are discussed in a concluding section.

\section{(Conformal) Aristotelian  geometries}

In this section we consider both non-conformal and conformal Aristotelian geometries, which we discuss separately, starting with the non-conformal case.

\subsection{$p$-brane Aristotelian geometry}

In this paper, we consider $p$-brane extensions \cite{Bergshoeff:2023rkk} of (particle) Aristotelian geometries that appeared in \cite{Figueroa-OFarrill:2018ilb,Figueroa-OFarrill:2020gpr,Bidussi:2021nmp,Jain:2021ibh}. We define a $p$-brane Aristotelian $G$-structure on a $D$-dimensional manifold as the intersection of a Galilean and a Carrollian $G$-structure; its structure group $G$ is thus given by:
\begin{align}
  \label{eq:structgroupSNC}
   G =  \mathrm{SO}(p,1) \times \mathrm{SO}(D-p-1) \,,
\end{align}
for integer $0 \le p \le D-2$. We will refer to the geometry with this local structure group as `$p$-brane Aristotelian geometry'.
We will call the $p+1$ directions on which $ \mathrm{SO}(p,1)$ acts the `longitudinal' directions and the $D-p-1$ directions on which $ \mathrm{SO}(D-p-1)$ acts the `transversal' directions. Likewise, we will often refer to the SO$(p,1)$ and SO$(D-p-1)$ factors of the structure group as the `longitudinal Lorentz transformations' and the `transversal rotations'.

To describe the geometry, we introduce a  `longitudinal Vielbein' $\tau_\mu{}^A$ ($A=0,1, \cdots ,p$) and a `transversal Vielbein' $e_\mu{}^{a}$ ($a = p+1, \cdots, D-1$). The flat longitudinal index $A$ will be freely raised and lowered with a $(p+1)$-dimensional Minkowski metric $\eta_{AB} = \mathrm{diag}(-1,1,\cdots ,1)$, whereas for the flat transversal index $a$ this will be done using a $(D-p-1)$-dimensional Euclidean metric $\delta_{ab}$. These one-forms transform under the structure group \eqref{eq:structgroupSNC} according to the following local transformation rules:
\begin{align} \label{eq:localtrafosframeSNC}
  \delta \tau_\mu{}^A &=  \lambda^A{}_B \tau_\mu{}^B \,, \qquad \qquad \qquad \delta e_\mu{}^{a} = \lambda^{a}{}_{b} e_\mu{}^{b}\,.
\end{align}
Here, $\lambda^{AB} = -\lambda^{BA}$ corresponds to the parameters of longitudinal SO$(p,1)$ Lorentz transformations and  $\lambda^{ab} = - \lambda^{ba}$ to that of transversal SO$(D-p-1)$ rotations. We also introduce  an `inverse longitudinal Vielbein' $\tau_A{}^\mu$ and an `inverse transversal Vielbein' $e_{a}{}^\mu$  as the vectors dual to $\tau_{\mu}{}^{A}$ and $e_{\mu}{}^{a}$; i.e., one has the following relations:
\begin{alignat}{3}
  \label{eq:invVielbeineSNC}
  & \tau_A{}^\mu \tau_\mu{}^B = \delta_A^B \,, \qquad \qquad \qquad & & \tau_A{}^\mu e_\mu{}^{a} = 0 \,, \qquad \qquad \qquad \qquad & & e_{a}{}^\mu \tau_\mu{}^A = 0 \,, \nonumber \\
  & e_\mu{}^{a} e_{b}{}^\mu = \delta_{b}^{a} \,, \qquad \qquad \qquad & & \tau_\mu{}^A \tau_A{}^\nu + e_\mu{}^{a} e_{a}{}^\nu = \delta_\mu^\nu \,.
\end{alignat}

A structure group connection $\Omega_\mu$ is introduced as the following one-form that takes values in the Lie algebra of \eqref{eq:structgroupSNC}:
\begin{equation}
  \Omega_\mu = \frac12 \omega_\mu{}^{AB} M_{AB} + \frac12 \omega_\mu{}^{ab} J_{ab} \,,
\end{equation}
where $M_{AB}=-M_{BA}$ and  $J_{ab} = -J_{ba}$ are generators of the Lie algebras of SO$(p,1)$ and  SO$(D-p-1)$, respectively. We will refer to $\omega_\mu{}^{AB}= -\omega_\mu{}^{BA}$ and  $\omega_\mu{}^{ab} = - \omega_\mu{}^{ba}$  as spin-connections for longitudinal Lorentz transformations and transversal rotations, respectively. They transform as follows under infinitesimal local SO$(p,1)$ and SO$(D-p-1)$ transformations:
\begin{align}
  \label{eq:localtrafosdepSNC}
    \delta \omega_\mu{}^{AB} &= \partial_\mu \lambda^{AB} + 2 \lambda^{[A|C|} \omega_{\mu C}{}^{B]}\,, \qquad \qquad
    \delta \omega_\mu{}^{ab} = \partial_\mu \lambda^{ab} + 2 \lambda^{[a|c|} \omega_{\mu c}{}^{b]} \,.
\end{align}
The curvatures of these spin-connection fields are given by:
\begin{align}
\label{eq:OmegaCurvatures}
R_{\mu\nu}{}^{AB}(M) \ &\equiv \ 2\partial_{[\mu}\omega_{\nu]}{}^{AB} -2 \omega_{[\mu}{}^{AC}\omega_{\nu] C}{}^B\,,\\[.1truecm]
R_{\mu\nu}{}^{ab}(J) \ &\equiv \ 2\partial_{[\mu}\omega_{\nu]}{}^{ab} -2 \omega_{[\mu}{}^{ac}\omega_{\nu] c}{}^b\,.
\end{align}
We also define their torsion tensors as
\begin{align}
  T_{\mu\nu}{}^A\ &\equiv \ 2 \partial_{[\mu} \tau_{\nu]}{}^A - 2 \omega_{[\mu}{}^{AB} \tau_{\nu] B} \,, \label{eq:torsionVielbSNC1} \\  E_{\mu\nu}{}^{a}\ &\equiv \ 2 \partial_{[\mu} e_{\nu]}{}^{a} - 2 \omega_{[\mu}{}^{ab} e_{\nu] b}  \label{eq:torsionVielbSNC2} \,.
\end{align}
The local infinitesimal SO$(p,1)$ and SO$(D-p-1)$ transformations of $T_{\mu\nu}{}^{A}$ and $E_{\mu\nu}{}^{a}$ are given by:
\begin{align}
  \label{eq:trafoTATAp}
  \delta T_{\mu\nu}{}^A = \lambda^A{}_B T_{\mu\nu}{}^B \,, \qquad \qquad \qquad \delta E_{\mu\nu}{}^{a} = \lambda^{a}{}_{b} E_{\mu\nu}{}^{b}\,.
\end{align}

For the purpose of classifying Aristotelian geometries, we are interested in finding the answers to the following two questions \cite{Figueroa-OFarrill:2020gpr,Bergshoeff:2023rkk}.
\vskip .15truecm

\noindent (i) Which components of the spin-connections $\omega_{\mu}{}^{AB}$ and $\omega_{\mu}{}^{ab}$ do not occur in any of the components of the torsion tensors $T_{\mu\nu}{}^{A}$ and $E_{\mu\nu}{}^{a}$? Such spin-connection components cannot be solved for in terms of the Vielbein fields and torsion tensor components. They remain independent.
\vskip .15truecm

\noindent (ii) Which components of the torsion tensors $T_{\mu\nu}{}^{A}$ and $E_{\mu\nu}{}^{a}$ do not contain any of the components of the spin-connections $\omega_{\mu}{}^{AB}$ and $\omega_{\mu}{}^{ab}$? Such torsion tensor components are called intrinsic. Setting them to zero leads to constraints on the geometry. They therefore play a crucial role in classifying the different Aristotelian geometries.
\vskip .3truecm

\noindent The answer to (i) can be obtained by examining the system of equations
\begin{align}
\label{eq:spinconnections11}
 2 \omega_{[\mu}{}^{AB} \tau_{\nu]B} = 0 \,, \qquad \qquad \qquad 2 \omega_{[\mu}{}^{ab} e_{\nu]b} = 0 \,,
\end{align}
that is obtained by setting to zero those terms in the definitions of $T_{\mu\nu}{}^{A}$ and $E_{\mu\nu}{}^{a}$ that depend on the two spin-connections, see the expressions \eqref{eq:torsionVielbSNC1} and \eqref{eq:torsionVielbSNC2}. The independent spin-connection components then correspond to the non-trivial solutions of this system. To distinguish the different spin-connection and torsion components, we decompose any curved index $\mu$ into longitudinal and transversal indices $A$ and $a$, according to the following decomposition rule for an arbitrary one-form $V_\mu$:
\begin{equation}\label{decomposition}
V_\mu = \tau_\mu{}^A V_A + e_\mu{}^a V_a\hskip .9truecm \textrm{or}\hskip .9truecm V_A = \tau_A{}^\mu V_\mu\ \ \textrm{and}\ \ V_a = e_a{}^\mu V_\mu\,.
\end{equation}
The same decomposition rule will be applied if the index $\mu$ is carried by a tensor. Using this rule, one finds that the system of equations \eqref{eq:spinconnections11} is equivalent to:
\begin{align}
\label{eq:spinconnections1}
\omega_{[A}{}^C{}_{B]} = 0\,, \ \ \ \omega_a{}^{BC}=0\,, \hskip 2truecm
\omega_{[a}{}^c{}_{b]} = 0\,, \ \ \ \omega_A{}^{ab}=0\,.
\end{align}
These equations only have the trivial solution
\begin{equation}\label{eq:solindepspinconnections}
\omega_\mu{}^{AB} = \omega_\mu{}^{ab}=0\,,
\end{equation}
so that there are no independent spin-connection fields that cannot be solved for in terms of the Vielbein fields and torsion tensor components.
To find the answer to the second question (ii) one scans the different components of the definitions of $T_{\mu\nu}{}^{A}$ and $E_{\mu\nu}{}^{a}$ and checks whether spin-connection components drop out. In this way one finds the following intrinsic torsion tensors:\,\footnote{We note that a further fine-tuning occurs when the longitudinal or transverse representations allow for self-duality conditions. Such situations occur when taking a particle in five spacetime dimensions where the transversal SO(4) allows self-duality conditions or when taking a string, where the longitudinal SO(1,1) allows self-duality conditions.}
\begin{equation}\label{intrinsic1}
T_{ab}{}^A\,,\hskip .5truecm T_a{}^{\{AB\}}\,,\hskip .5truecm T_a{}^A{}_A\,,\hskip .5truecm E_{AB}{}^a\,,\hskip .5truecm E_A{}^{\{ab\}}\,,\hskip .5truecm
E_A{}^a{}_a\,.
\end{equation}
We use here a notation where $\{AB\}$ indicates the symmetric traceless part of $AB$. The same applies to $\{ab\}$.

The remaining torsion tensor components all contain a spin-connection field. They are given by
\begin{equation}\label{conventional}
T_a{}^{[AB]}\,, \hskip .5truecm T_{AB}{}^C\,,\hskip .5truecm E_A{}^{[ab]}\,,\hskip .5truecm E_{ab}{}^{c}\,.
\end{equation}
Note that the spin-connection terms in the expressions \eqref{eq:torsionVielbSNC1} and \eqref{eq:torsionVielbSNC2} for these torsion components are multiplied by Vielbeine. As a consequence, one can invert these expressions to write the dependent spin-connection fields in terms of Vielbeine and the torsion tensor components \eqref{conventional}. Setting these torsion tensor components to zero therefore does not lead to constraints on the geometry but rather amounts to choosing a specific connection. Following the supergravity literature, we will call the components \eqref{conventional} conventional torsion tensor components.

In the case at hand we find that the number of conventional torsion tensor components equals the number of spin-connection components.  This is consistent with the fact that we found above that there are no independent spin-connections. Setting all conventional torsion tensor components given in eq.~\eqref{conventional} to zero,  we find the following expressions for the different spin-connection components in terms of the Vielbeine:
\begin{subequations}
\label{depspincnconf}
\begin{align}
\omega_{A,BC}(\tau,e) &= \frac{1}{2} \tau_{BC,A} - \tau_{A[B,C]} \,,\label{solutions1}  \\
\omega_{a,AB}(\tau,e) &=  -\tau_{a[A,B]}\,,\label{solutions2} \\
\omega_{a,bc}(\tau,e) &=  \frac{1}{2} e_{bc,a} - e_{a[b,c]}\,,\label{solutions3} \\
\omega_{A,ab}(\tau,e) &=   - e_{A[a,b]}\,,\label{solutions4}
\end{align}
\end{subequations}
where we have put a comma in $\omega_{A,BC} = -\omega_{A,CB}$ to distinguish between the first index and the second anti-symmetric pair of indices.\,\footnote{Here and below we will only put a comma in cases that confusion could arise.} Furthermore, we have defined
\begin{equation}
\label{eq:deftauemunu}
\tau_{\mu \nu}{}^A \equiv 2  \,\partial_{[\mu} \tau_{\nu]}{}^A \hskip .5 truecm \textrm{and}\hskip .5truecm
e_{\mu \nu}{}^a \equiv 2  \,\partial_{[\mu} e_{\nu]}{}^a\,.
\end{equation}
We stress that the dependent spin-connections are only solved as in \eqref{depspincnconf}, if we equate the full set \eqref{conventional} of conventional torsion tensor components to zero. The dependent spin-connection components \eqref{depspincnconf} still transform as the independent ones given in \eqref{eq:localtrafosdepSNC} due to the fact that the constraints defined by setting the components \eqref{conventional} equal to zero are invariant under the Aristotelian structure group \eqref{eq:structgroupSNC}.

We are now ready to perform the classification of the different  $p$-brane Aristotelian geometries. We will classify the representations of the intrinsic torsion tensors since setting them to zero gives rise to geometric constraints. In contrast to the Galilean and Carroll cases, the different intrinsic torsion tensor components given in eq.~\eqref{intrinsic1} are not connected to each other by boost transformations. Therefore, each of these six intrinsic torsion tensor components can separately be set equal to zero.  For the general $p$-brane case, $ 0 < p < D-2 $,  this leads to
\begin{equation}
\sum_{q=0}^6
\begin{pmatrix}6\cr q
\end{pmatrix} = 64
\end{equation}
different Aristotelian geometries. The cases of particle ($p=0$) and domain wall ($p=D-2$) are special since then 2 of the 6 intrinsic torsion tensors vanish. For instance, in the particle case one ends up with the following 4 intrinsic torsion tensors
\begin{equation}\label{intrinsic2}
T_{ab}{}^0\,,\hskip .5truecm T_a{}^0{}_0\,,\hskip .5truecm  E_0{}^{\{ab\}}\,,\hskip .5truecm
E_0{}^a{}_a\,,
\end{equation}
where we have indicated the single time direction $A$ with $A=0$. Therefore, the particle case leads to only
\begin{equation}
\sum_{q=0}^4
\begin{pmatrix}4\cr q
\end{pmatrix} = 16
\end{equation}
different Aristotelian geometries.  The geometric interpretation of each of these 16 geometric constraints can be found in \cite{Figueroa-OFarrill:2020gpr}. Similarly, in the domain wall case, indicating the single flat transverse direction $a$ with $a=y$,  one obtains the following 4 intrinsic torsion tensors:
\begin{equation}\label{intrinsic3}
T_y{}^{\{AB\}}\,,\hskip .5truecm T_y{}^A{}_A\,,\hskip .5truecm E_{AB}{}^y\,,\hskip .5truecm E_A{}^y{}_y\,,
\end{equation}
that as in the particle case gives rise to 16 different Aristotelian geometries.

\subsection{Conformal extensions} \label{ssec:confgeom}

In this section, we provide a conformal extension of the Aristotelian symmetries and geometry of the previous subsection. This will be used in section \ref{sec:argrav} to construct Aristotelian gravity theories via conformal calculus techniques.  We will discuss {\sl two} such conformal extensions: one that has two isotropic dilatations and that we will call the isotropic conformal Aristotelian algebra and a subalgebra which has a single an-isotropic dilatation that we will call the an-isotropic Aristotelian algebra.
\vskip .2truecm

\noindent {\bf The isotropic conformal Aristotelian algebra:}

\noindent One way to extend the Aristotelian symmetries \eqref{eq:structgroupSNC} with conformal transformations is by considering a direct sum of two conformal algebras:
\begin{align}
  \label{eq:galg}
  \mathfrak{so}(2,p+1) \oplus \mathfrak{so}(1,D-p) \,.
\end{align}
The algebra $\mathfrak{so}(2,p+1)$ is the conformal algebra in $p+1$ dimensions with Minkowski signature. We will refer to and denote its generators by:\,\footnote{The indices $A$ and $a$ used in this subsection take values in the same ranges as in the previous subsection.}
\begin{align}
  \label{eq:conf1gens}
  & \text{longitudinal translations} \ P_{A} \,, \qquad \qquad \qquad \text{longitudinal Lorentz transformations} \ M_{AB} \,, \nonumber \\ & \text{longitudinal special conformal transformations} \ K_{A} \,, \qquad
  \text{the longitudinal dilatation} \ D_1 \,.
\end{align}
The non-zero commutation relations of $\mathfrak{so}(2,p+1)$ are given by
\begin{alignat}{2}
  \label{eq:conf1comms}
  \comm{M_{AB}}{M_{CD}} &= - 4 \eta_{[A[C} M_{D]B]} \,, \qquad \qquad & \comm{P_{A}}{M_{BC}} &= - 2 \eta_{A[B} P_{C]} \,, \nonumber \\
  \comm{K_{A}}{M_{BC}} &= - 2 \eta_{A[B} K_{C]} \,, \qquad \qquad & \comm{P_{A}}{K_{B}} &= 2 \eta_{AB} D_1 + 2 M_{AB} \,, \nonumber \\
  \comm{D_1}{P_{A}} &= - P_{A} \,, \qquad \qquad & \comm{D_1}{K_{A}} &= K_{A} \,.
\end{alignat}
The algebra $\mathfrak{so}(1,D-p)$ on the other hand corresponds to the conformal algebra in $D-p-1$ Euclidean dimensions and we will refer to and denote its generators by
\begin{align}
  \label{eq:conf2gens}
  & \text{transversal translations} \ P_{a} \,, \qquad \qquad \qquad \qquad \qquad \qquad \text{transversal rotations} \ J_{ab} \,, \nonumber \\ & \text{transversal special conformal transformations} \ K_{a} \,, \qquad \text{the transversal dilatation} \ D_2 \,.
\end{align}
The commutation relations of $\mathfrak{so}(1,D-p)$ are obtained from \eqref{eq:conf1comms}, by replacing $\eta_{AB}$ with $\delta_{ab}$ and $P_{A}$, $M_{AB}$, $K_{A}$ and $D_1$ with $P_{a}$, $J_{ab}$, $K_{a}$ and $D_2$ respectively. The direct sum in \eqref{eq:galg} is interpreted as a direct sum of Lie algebras, i.e., the algebras $\mathfrak{so}(2,p+1)$ and $\mathfrak{so}(1,D-p)$ commute with each other.

To describe our conformal extension of Aristotelian geometry, we supplement the longitudinal and transversal Vielbeine $\tau_{\mu}{}^{A}$, $e_{\mu}{}^{a}$ and the spin-connections $\omega_{\mu}{}^{AB}$, $\omega_{\mu}{}^{ab}$ of the previous subsection with four extra fields $f_{\mu}{}^{A}$, $f_{\mu}{}^{a}$, $b_{\mu}$ and $c_{\mu}$. We will call the field $f_{\mu}{}^{A}$, resp. $f_{\mu}{}^{a}$ the longitudinal, resp. transversal special conformal gauge field, and $b_{\mu}$, resp. $c_{\mu}$ the longitudinal, resp. transversal dilatation gauge field. This terminology derives from the fact that we view the full set of fields as gauge fields that are associated to the generators of the algebra \eqref{eq:galg} as follows:
\begin{alignat}{4}
  \label{eq:argaugefieldid}
   \tau_{\mu}{}^{A} \ \ & \rightarrow \ \ P_{A} \,, \qquad & \omega_{\mu}{}^{AB} \ \ & \rightarrow \ \ M_{AB} \,, \qquad & f_{\mu}{}^{A} \ \ & \rightarrow \ \ K_{A} \,, \qquad & b_{\mu} \ \ & \rightarrow \ \ D_1 \,, \nonumber \\
   e_{\mu}{}^{a} \ \ & \rightarrow \ \ P_{a} \,, \qquad & \omega_{\mu}{}^{ab} \ \ & \rightarrow \ \ J_{ab} \,, \qquad & f_{\mu}{}^{a} \ \ & \rightarrow \ \ K_{a} \,, \qquad & c_{\mu} \ \ & \rightarrow \ \ D_2 \,.
\end{alignat}
We will mostly focus on the transformations generated by $M_{AB}$, $K_{A}$, $D_1$, $J_{ab}$, $K_{a}$ and $D_2$ that we will collectively call ``homogeneous conformal transformations''. Under the identification \eqref{eq:argaugefieldid} of the fields as gauge fields of the algebra \eqref{eq:galg}, one has the following gauge transformation rules under these homogeneous conformal transformations:
\begin{subequations}
  \label{eq:haconfrules0}
\begin{align}
  \delta \tau_{\mu}{}^{A} &= \lambda^{A}{}_{B} \tau_{\mu}{}^{B} + \sigma_1 \tau_{\mu}{}^{A} \,, \label{eq:taurule} \\
  \delta \omega_{\mu}{}^{AB} &= \partial_{\mu} \lambda^{AB} + 2 \lambda^{[A|C|} \omega_{\mu C}{}^{B]} - 4 \lambda_{K}{}^{[A} \tau_{\mu}{}^{B]} \,, \label{eq:omABrule0} \\
  \delta f_{\mu}{}^{A} &= \partial_{\mu} \lambda_{K}{}^{A} - \omega_{\mu}{}^{AB} \lambda_{K B} + b_{\mu} \lambda_{K}{}^{A} + \lambda^{A}{}_{B} f_{\mu}{}^{B} - \sigma_1 f_{\mu}{}^{A} \,, \label{eq:fArule0} \\
  \delta b_{\mu} &= \partial_{\mu} \sigma_1 + 2 \lambda_{K}{}^{A} \tau_{\mu A} \,, \label{eq:brule0} \\
  \delta e_{\mu}{}^{a} &=  \lambda^{a}{}_{b} e_{\mu}{}^{b} + \sigma_2 e_{\mu}{}^{a} \,, \label{eq:erule} \\
  \delta \omega_{\mu}{}^{ab} &= \partial_{\mu} \lambda^{ab} + 2 \lambda^{[a|c|} \omega_{\mu c}{}^{b]} - 4 \lambda_{K}{}^{[a} e_{\mu}{}^{b]} \,, \label{eq:omabrule0} \\
  \delta f_{\mu}{}^{a} &= \partial_{\mu} \lambda_{K}{}^{a} - \omega_{\mu}{}^{ab} \lambda_{K b} + c_{\mu} \lambda_{K}{}^{a} + \lambda^{a}{}_{b} f_{\mu}{}^{b}  - \sigma_2 f_{\mu}{}^{a} \,, \label{eq:farule0} \\
  \delta c_{\mu} &= \partial_{\mu} \sigma_2 + 2 \lambda_{K}{}^{a} e_{\mu a} \,, \label{eq:crule0}
\end{align}
\end{subequations}
where $\lambda^{AB}$, $\lambda_K{}^{A}$, $\sigma_1$, $\lambda^{ab}$, $\lambda_K{}^{a}$ and $\sigma_2$ are the parameters of infinitesimal $M_{AB}$, $K_{A}$, $D_1$, $J_{ab}$, $K_{a}$ and $D_2$ transformations, respectively.

Thus far, we have treated all gauge fields $\{\tau_{\mu}{}^{A}, \omega_{\mu}{}^{AB}, f_{\mu}{}^{A}, b_{\mu}, e_{\mu}{}^{a}, \omega_{\mu}{}^{ab}, f_{\mu}{}^{a}, c_{\mu}\}$ as independent. It is possible to turn a subset of these gauge fields into dependent ones and in this way obtain a representation of the homogeneous conformal transformations on a smaller number of independent fields. To do this, we first generalize the torsion tensors of \eqref{eq:torsionVielbSNC1} and \eqref{eq:torsionVielbSNC2} to
\begin{subequations}
  \label{eq:conftorsion}
\begin{align}
    \mathcal{T}_{\mu\nu}{}^{A} &\equiv 2 \partial_{[\mu} \tau_{\nu]}{}^{A} - 2 \omega_{[\mu}{}^{AB} \tau_{\nu]B} - 2 b_{[\mu} \tau_{\nu]}{}^{A} = T_{\mu\nu}{}^{A} - 2 b_{[\mu} \tau_{\nu]}{}^A \,, \label{eq:conftorsion1} \\
    \mathcal{E}_{\mu\nu}{}^{a} &\equiv 2 \partial_{[\mu} e_{\nu]}{}^{a} - 2 \omega_{[\mu}{}^{ab} e_{\nu]b} - 2 c_{[\mu} e_{\nu]}{}^{a} = E_{\mu\nu}{}^{a} - 2 c_{[\mu} e_{\nu]}{}^{a} \,. \label{eq:conftorsion2}
\end{align}
\end{subequations}
These torsion tensors transform covariantly under the transformations \eqref{eq:haconfrules0}:
\begin{align}
  \delta \mathcal{T}_{\mu\nu}{}^{A} &=  \lambda^{A}{}_{B} \mathcal{T}_{\mu\nu}{}^{B} + \sigma_1 \mathcal{T}_{\mu\nu}{}^{A} \,, \qquad \qquad \delta \mathcal{E}_{\mu\nu}{}^{a} =  \lambda^{a}{}_{b} \mathcal{E}_{\mu\nu}{}^{b} + \sigma_2 \mathcal{E}_{\mu\nu}{}^{a} \,.
\end{align}

Inspection of the definitions \eqref{eq:conftorsion} shows that the following torsion tensor components
\begin{align}
  \label{eq:confintrtorsion}
  \mathcal{T}_{ab}{}^{A} \,, \quad \mathcal{T}_{a}{}^{\{A,B\}} \,, \qquad \qquad \mathcal{E}_{AB}{}^{a} \,, \quad \mathcal{E}_{A}{}^{\{a,b\}} \,,
\end{align}
do not contain any components of the spin-connections and dilatation gauge fields $\omega_{\mu}{}^{AB}$, $\omega_{\mu}{}^{ab}$, $b_{\mu}$ and $c_{\mu}$. The components \eqref{eq:confintrtorsion} thus constitute a conformal generalization of the notion of intrinsic torsion, introduced in the previous subsection. Setting the remaining, non-intrinsic torsion tensor components equal to zero in \eqref{eq:conftorsion} gives a set of equations that can be solved for the spin-connections $\omega_{\mu}{}^{AB}$, $\omega_{\mu}{}^{ab}$ and the dilatation gauge field components $b_{a}$ and $c_{A}$. Explicitly, one finds the following expressions for these fields:
\begin{align}
  \label{eq:confdepombc}
  \omega_{\mu}{}^{AB} &= \tau_{\mu}{}^{C} \left(\frac12 \tau^{AB}{}_{,C} -\tau_{C}{}^{[A,B]} - 2 b^{[A} \delta^{B]}_{C} \right) - e_{\mu}{}^{a} \tau_{a}{}^{[A,B]} \,, \nonumber \\
  \omega_{\mu}{}^{ab} &= - \tau_{\mu}{}^{A} e_{A}{}^{[a,b]} + e_{\mu}{}^{c} \left(\frac12 e^{ab}{}_{,c} - e_{c}{}^{[a,b]} - 2 c^{[a} \delta^{b]}_{c} \right) \,,  \\
  b_{a} &= \frac{1}{(p+1)} T_{aA}{}^{A} \left(= \frac{1}{(p+1)} \tau_{a A}{}^{A}\right) \,, \quad c_{A} = \frac{1}{(D-p-1)} E_{Aa}{}^{a} \left(= \frac{1}{(D-p-1)} e_{A a}{}^{a} \right)\,,\nonumber
\end{align}
where we used the notation \eqref{eq:deftauemunu}. In what follows, we will always assume that $\omega_{\mu}{}^{AB}$, $\omega_{\mu}{}^{ab}$, $b_{a}$ and $c_{A}$ are given by these expressions, wherever they occur.

For $p \neq 0$ and $p \neq 1$, we can also turn the special conformal gauge field $f_{\mu}{}^{A}$ into a dependent field, by solving it from the conventional constraint:
\begin{align}
  \label{eq:RJABconstr}
  \mathcal{R}_{\mu B}{}^{AB}(M) = 0 \,, \qquad \text{where} \ \ \mathcal{R}_{\mu\nu}{}^{AB}(M) \equiv 2 \partial_{[\mu} \omega_{\nu]}{}^{AB} - 2 \omega_{[\mu}{}^{[A|C|} \omega_{\nu]C}{}^{B]} + 8 f_{[\mu}{}^{[A} \tau_{\nu]}{}^{B]} \,.
\end{align}
The curvature $\mathcal{R}_{\mu\nu}{}^{AB}(M)$ thus corresponds to the conformally covariant field strength of the $M_{AB}$ transformations. The solution for $f_{\mu}{}^{A}$ that is found from the constraint \eqref{eq:RJABconstr} is given by
\begin{align}
  \label{eq:genpfAsol}
  f_{\mu}{}^{A} &= -\frac{1}{2 p (p-1)} \left[p R_{\mu B}{}^{AB}(M) - e_{\mu}{}^{a} R_{aB}{}^{AB}(M) - \frac12 \tau_{\mu}{}^{A} R_{BC}{}^{BC}(M)\right] \,,
\end{align}
where $R_{\mu\nu}{}^{AB}(M)$ is defined as in \eqref{eq:OmegaCurvatures} (and is thus given by $\mathcal{R}_{\mu\nu}{}^{AB}(M)$ without the term involving $f_{\mu}{}^{A}$).

The case $p=1$ requires a special treatment.\,\footnote{Note that for $p=0$ there is no curvature for longitudinal rotations to start with.} When $p=1$, one can still impose \eqref{eq:RJABconstr}, but these constraints no longer allow one to solve for all components of $f_{\mu}{}^{A}$. This is because for $p=1$, the constraints \eqref{eq:RJABconstr} can be equivalently written as\,\footnote{This is because for $p=1$, one can write
  \begin{align}
    \mathcal{R}_{BC}{}^{AC}(M) \propto \varepsilon_{BC} \varepsilon^{DE} \varepsilon^{AC} \varepsilon_{FG} \mathcal{R}_{DE}{}^{FG}(M) \propto \delta_B^A \mathcal{R}_{CD}{}^{CD}(M) \,.
  \end{align}
}:
\begin{align}
  \label{eq:p1fAconstr}
  \mathcal{R}_{aB}{}^{AB}(M) = 0 \,, \qquad \qquad \text{and} \qquad \qquad \mathcal{R}_{AB}{}^{AB}(M) = 0 \,.
\end{align}
The first of these equations can be solved for $f_a{}^{A}$, while the second can be solved for $f_{A}{}^{A}$:
\begin{align}
  f_a{}^{A} = -\frac{1}{2} R_{aB}{}^{AB}(M) \,, \qquad \qquad f_{A}{}^{A} = - \frac{1}{4} R_{AB}{}^{AB}(M) \,.
\end{align}
Note in particular that it is not possible to solve for the traceless part $f_{B}{}^{A} - \tfrac{1}{2} \delta_B^A f_{C}{}^{C}$.

For $p \neq D-2$ and $p \neq D-3$, the second special conformal gauge field $f_{\mu}{}^{a}$ can likewise be solved from the constraints:
\begin{align}
  \label{eq:genpfaconstr}
  \mathcal{R}_{\mu b}{}^{ab}(J) = 0  \qquad \text{where} \ \ \mathcal{R}_{\mu\nu}{}^{ab}(J) \equiv 2 \partial_{[\mu} \omega_{\nu]}{}^{ab} - 2 \omega_{[\mu}{}^{[a|c|} \omega_{\nu]c}{}^{b]} + 8 f_{[\mu}{}^{[a} e_{\nu]}{}^{b]} \,.
\end{align}
One finds:
\begin{align} \label{eq:genpfasol}
  f_{\mu}{}^{a} &= -\frac{1}{2(D-p-3)(D-p-2)} \left[ (D-p-2) R_{\mu b}{}^{ab}(J) - \tau_{\mu}{}^{A} R_{Ab}{}^{ab}(J) - \tfrac12 e_{\mu}{}^{a} R_{bc}{}^{bc}(J)\right] \,,
\end{align}
where $R_{\mu\nu}{}^{ab}(J)$ is defined in \eqref{eq:OmegaCurvatures} (and thus corresponds to $\mathcal{R}_{\mu\nu}{}^{ab}(J)$ without the term involving $f_{\mu}{}^{a}$).

When $p = D-3$\,\footnote{For $p=D-2$ there is no curvature for transverse rotations to start with.}, the constraints \eqref{eq:genpfaconstr} are equivalent to
\begin{align}
  \label{eq:p1faconstr}
  \mathcal{R}_{Ab}{}^{ab}(J) = 0 \,, \qquad \qquad \text{and} \qquad \qquad \mathcal{R}_{ab}{}^{ab}(J) = 0 \,,
\end{align}
and can only be used to turn $f_{A}{}^{a}$ and the trace $f_{a}{}^{a}$ into dependent fields:
\begin{align}
  f_{A}{}^{a} &= - \frac{1}{2} R_{Ab}{}^{ab}(J) \,, \qquad \qquad f_{a}{}^{a} = - \frac{1}{4} R_{ab}{}^{ab}(J) \,.
\end{align}
Like what happened for $f_{\mu}{}^{A}$ in case $p=1$, one can not solve for the traceless part $f_{b}{}^{a} - \tfrac{1}{2} \delta_{b}^{a} f_{c}{}^{c}$. 

This finishes our discussion of the isotropic conformal Aristotelian algebra.

\vskip .2truecm

\noindent {\bf The an-isotropic conformal Aristotelian algebra:}

\noindent The algebra \eqref{eq:galg} that includes two dilatations is not the only way of providing a conformal extension of the Aristotelian symmetries. It is also possible to extend the Aristotelian symmetries with a single an-isotropic dilatation. Such an extension corresponds to the subalgebra of \eqref{eq:galg} spanned by the generators
\begin{align}
    \label{eq:galganiso0}
 \{M_{AB}, J_{ab}, P_{A}, P_{a}, D \equiv z D_1 + D_2 \} \,,
\end{align}
where the critical exponent $z \in \mathbb{R}$ parametrizes the manner in which the longitudinal directions scale differently from the transversal ones under the dilatation $D$. From the commutation relations of \eqref{eq:galg}, one finds that those of the subalgebra \eqref{eq:galganiso0} are given by
\begin{alignat}{2}
    \label{eq:galganiso}
    \comm{M_{AB}}{M_{CD}} &= - 4 \eta_{[A[C} M_{D]B]} \,, \qquad \qquad & \comm{P_{A}}{M_{BC}} &= - 2 \eta_{A[B} P_{C]} \,, \nonumber \\
    \comm{J_{ab}}{J_{cd}} &= - 4 \delta_{[a[c} J_{d]b]} \,, \qquad \qquad & \comm{P_{a}}{J_{bc}} &= - 2 \delta_{a[b} P_{c]} \,, \nonumber \\
    \comm{D}{P_{A}} &= - z P_{A} \,, \qquad \qquad & \comm{D}{P_{a}} &= - P_{a} \,.
\end{alignat}
As for the isotropic algebra, we view the Vielbeine $\tau_{\mu}{}^{A}$, resp. $e_{\mu}{}^{a}$ as gauge fields associated to $P_{A}$, resp. $P_{a}$ and the spin-connections $\omega_{\mu}{}^{AB}$, $\omega_{\mu}{}^{ab}$ as gauge fields for $M_{AB}$, resp. $J_{ab}$. We furthermore also introduce a gauge field $b_{\mu}$, associated to the dilatation $D$. The algebra \eqref{eq:galganiso} then implies that the fields $\tau_{\mu}{}^{A}$, $e_{\mu}{}^{a}$, $\omega_{\mu}{}^{AB}$, $\omega_{\mu}{}^{ab}$ and $b_{\mu}$ transform as follows under the transformations generated by $M_{AB}$, $J_{ab}$ and $D$ (with the respective parameters $\lambda^{AB}$, $\lambda^{ab}$ and $\lambda_{D}$):
\begin{align} \label{eq:localtrafosTorsionsDilatation1}
    \delta \tau_{\mu}{}^{A} &= \lambda^{A}{}_{B} \tau_{\mu}{}^{B} + z \lambda_{D} \tau_{\mu}{}^{A} \,, \qquad \quad
    \delta e_{\mu}{}^{a} = \lambda^{a}{}_{b} e_{\mu}{}^{b} + \lambda_{D} e_{\mu}{}^{a} \,, \qquad \quad \delta b_{\mu} = \partial_{\mu} \lambda_{D} \,, \nonumber \\
    \delta \omega_{\mu}{}^{AB} &= \partial_{\mu} \lambda^{AB} + 2 \lambda^{[A|C|} \omega_{\mu C}{}^{B]} \,, \qquad \qquad \quad
    \delta \omega_{\mu}{}^{ab} = \partial_{\mu} \lambda^{ab} + 2 \lambda^{[a|c|} \omega_{\mu c}{}^{b]} \,.
\end{align}
The appropriate generalization of the torsion tensors \eqref{eq:torsionVielbSNC1}, \eqref{eq:torsionVielbSNC2} is now given by:
\begin{subequations}
\begin{align}
\tilde{\mathcal{T}}_{\mu\nu}{}^A &= T_{\mu\nu}{}^A- 2z\, b_{[\mu} \tau_{\nu]}{}^A \,, \label{eq:torsionVielbSNC1Dilatation} \\  \tilde{\mathcal{E}}_{\mu\nu}{}^{a} &=  E_{\mu\nu}{}^{a}
- 2 b_{[\mu} e_{\nu]}{}^a\label{eq:torsionVielbSNC2Dilatation} \,,
\end{align}
\end{subequations}
since these transform covariantly under dilatations, as well as SO$(p,1)$ and SO$(D-p-1)$:
\begin{align} \label{eq:localtrafosTorsionsDilatation}
  \delta \tilde{\mathcal{T}}_{\mu\nu}{}^A &= \lambda^{A}{}_{B} \tilde{\mathcal{T}}_{\mu\nu}{}^{B} +
  z\,\lambda_D \, \tilde{\mathcal{T}}_{\mu\nu}{}^A
  \,, \qquad \qquad \delta \tilde{\mathcal{E}}_{\mu\nu}{}^a = \lambda^{a}{}_{b} \tilde{\mathcal{E}}_{\mu\nu}{}^{b} + \lambda_D\, \tilde{\mathcal{E}}_{\mu\nu}{}^a\,.
\end{align}

Examining the components of $\tilde{\mathcal{T}}_{\mu\nu}{}^{A}$ and $\tilde{\mathcal{E}}_{\mu\nu}{}^{a}$, one finds that the spin-connection and dilatation gauge fields drop out of the following torsion components:
\begin{equation}\label{intrinsic1Dilalation}
\tilde{\mathcal{T}}_{ab}{}^A\,,\hskip .5truecm \tilde{\mathcal{T}}_a{}^{\{A,B\}}\,,\hskip .5truecm \tilde{\mathcal{E}}_{AB}{}^a\,,\hskip .5truecm \tilde{\mathcal{E}}_A{}^{\{a,b\}}\,.
\end{equation}
These therefore constitute the intrinsic torsion tensor components. All the remaining torsion components depend on the spin-connection and dilatation gauge fields, and are given by
\begin{alignat}{3}\label{eq:conventialDilatation}
& \tilde{\mathcal{T}}_a{}^A{}_A\,,\hskip .9truecm & &
\tilde{\mathcal{T}}_{a}{}^{[A,B]} \ (= T_a{}^{[A,B]})\, \hskip .9truecm & & \tilde{\mathcal{T}}_{AB}{}^{C} \ (= T_{AB}{}^C)\,,\\ \nonumber
& \tilde{\mathcal{E}}_A{}^a{}_a\,,\hskip .9truecm & &
\tilde{\mathcal{E}}_{A}{}^{[a,b]} \ (= E_A{}^{[a,b]})\,,\hskip .9truecm & & \tilde{\mathcal{E}}_{ab}{}^{c} \ (= E_{ab}{}^{c}) \,,
\end{alignat}
where we have indicated in parentheses that some of these are equal to corresponding components of $T_{\mu\nu}{}^{A}$ and $E_{\mu\nu}{}^{a}$, since they do not contain dilatation gauge field components. Setting all the components \eqref{eq:conventialDilatation} to zero allows to completely solve the spin-connection and the dilatation gauge field in terms of the Vielbein fields. There are thus no undetermined spin-connection or dilatation gauge field components. The solutions for $\omega_{a,AB}(\tau)$ and $\omega_{A,ab}(e)$ are as given in \eqref{solutions2} and \eqref{solutions4}, while the other two spin-connections \eqref{solutions1} and \eqref{solutions3} get contributions from the dilatation gauge field. The spin-connection components in this case are given by:
\begin{eqnarray}\label{depspinc}
\omega_{A,BC}(\tau, e, b) &=& \omega_{A,BC}(\tau, e) -2z\,  b_{[B}  \eta_{C]A} \,,\label{solutionsconformal1}\\[.1truecm]
\omega_{a,AB}(\tau, e, b) &=&  \omega_{a,AB}(\tau, e)\,, \label{solutionsconformal2}\\[.1truecm]
\omega_{a,bc}(\tau, e, b) &=&  \omega_{a,bc}(\tau, e) -2\,  b_{[b}  \delta_{c]a}\,,\label{solutionsconformal3}\\[.1truecm]
\omega_{A,ab}(\tau, e, b) &=&   \omega_{A,ab}(\tau, e)\,. \label{solutionsconformal4}
\end{eqnarray}
The two constraints $\tilde{\mathcal{E}}_{A}{}^a{}_{a}=\, 0$ and $\tilde{\mathcal{T}}_{a}{}^A{}_{A}=\,0$ lead to the following expression for the dilatation gauge field components
\begin{align}
b_a =&\, \frac{1}{z(p+1)}{T}_{a}{}^A{}_{A} \left(= \frac{1}{z(p+1)}{\tau}_{a}{}^A{}_{A}\right) \,, \label{eq:Conformalb1} \\
b_A =&\, \frac{1}{D-p-1}{E}_{A}{}^a{}_{a} \left(= \frac{1}{D-p-1}{e}_{A}{}^a{}_{a}\right) \,. \label{eq:Conformalb2}
\end{align}
The spin-connection and the dilatation gauge fields transform according to the conformal homogeneous Aristotelian algebra transformations of \eqref{eq:localtrafosTorsionsDilatation1}. 

This finishes our discussion of the an-isotropic conformal Aristotelian algebra. 
Differently from the non-conformal case, in both the isotropic and an-isotropic conformal cases discussed above we have only four torsion components (\eqref{eq:confintrtorsion} or \eqref{intrinsic1Dilalation}) that we can independently set to zero to formulate different conformal Aristotelian geometric constraints. Note moreover that the intrinsic torsion components lie in both cases in the same $\mathrm{SO}(1,p) \times \mathrm{SO}(D-p-1)$ representations.  For the general $p$-brane case, $ 0 < p < D-2 $,  we then get
\begin{equation}
\sum_{q=0}^4
\begin{pmatrix}4\cr q
\end{pmatrix} = 16 
\end{equation}
different conformal Aristotelian geometries.
For the particle case $p=\,0$ the only remaining intrinsic torsion tensors are given by
\begin{equation}
\mathcal{T}_{ab}{}^A\,, \qquad \mathcal{E}_A{}^{\{ab \}}\,, \qquad \qquad \text{or} \qquad \qquad \tilde{\mathcal{T}}_{ab}{}^A\,, \qquad \tilde{\mathcal{E}}_A{}^{\{ab \}}\,,
\end{equation}
and this leads to $4$ independent particle conformal Aristotelian geometries. A similar situation occurs for the domain wall case $p=D-2$.

\section{$p$-brane Aristotelian gravity from the conformal approach} \label{sec:argrav}

In this section, we will construct gravitational actions that are invariant under local Aristotelian symmetries and that do not exhibit hidden (Galilei or Carroll) boost symmetries. Such actions cannot be described as the limit of a Lorentz-invariant theory. Instead, we will construct them via a conformal compensating mechanism. In particular, we will consider different kinds of actions for compensating scalar fields, that are invariant under local homogeneous conformal transformations, by virtue of their coupling to the independent and dependent gauge fields of the conformal Aristotelian geometries discussed in section \ref{ssec:confgeom}. Actions for Aristotelian gravity are then obtained by gauge-fixing superfluous dilatations and special conformal symmetries.

Our conformal approach will lead to three different types of Aristotelian gravity, that are distinguished by whether their actions are constructed out of intrinsic torsion and/or Aristotelian curvature tensors. We will refer to these three types as ``electric", ``magnetic" and ``electric-magnetic". The actions of electric theories only consist of terms that are quadratic in intrinsic torsion components, whereas those of magnetic theories are only constructed out of Aristotelian curvatures. The Lagrangians of theories of the mixed electric-magnetic type on the other hand contain both squares of intrinsic torsion components, as well as curvature terms. Our terminology is inspired by the recent Carroll gravity literature  \cite{Henneaux:2021yzg,Hansen:2021fxi,Figueroa-OFarrill:2022mcy,Bergshoeff:2022eog}, where the distinction between electric and magnetic Carroll gravity theories is made on similar grounds. 

For all three types, the corresponding actions take the form of an arbitrary linear combination of a term that is reminiscent of (electric or magnetic) Galilei gravity and one that is reminiscent of (electric or magnetic) Carroll gravity. Galilei or Carroll boosts will thus be broken by (at least) one term in the action, so that generically our Aristotelian gravity actions do not exhibit any hidden boost symmetries. This is consistent with the fact that the actions below do not belong to the class of currently known Galilei or Carroll gravity actions.

We will apply the conformal compensating technique, starting from the two conformal extensions of the Aristotelian symmetry algebra, discussed in subsection \ref{ssec:confgeom}. In subsection \ref{ssec:isoconfcalc}, we will construct actions for two compensating scalar fields, that are invariant under the homogeneous special conformal transformations of the isotropic conformal Aristotelian algebra \eqref{eq:galg}. The need for two compensating scalars stems from the fact that this algebra contains two independent dilatation symmetries, each of which needs to be fixed by setting a scalar equal to a constant value. We will show that in this case one can build three types of scalar field actions that are gauge-equivalent to Aristotelian gravity theories of the electric, magnetic and electric-magnetic types. In order to be able to retrieve the curvature terms that appear in the magnetic and electric-magnetic theories, it is crucial that the conformal algebra \eqref{eq:galg} contains special conformal transformations. Indeed, such terms appear via couplings to the dependent special conformal gauge fields, which are constructed out of Aristotelian curvatures (see eqs. \eqref{eq:genpfAsol}, \eqref{eq:genpfasol}). The an-isotropic conformal algebra \eqref{eq:galganiso} only contains one dilatation. In using the conformal approach with this algebra, it thus suffices to construct actions for a single compensating scalar. Moreover, since \eqref{eq:galganiso} does not contain any special conformal transformations, only theories of the purely electric type can be constructed. We will discuss this case in subsection \ref{ssec:anisoconfcalc}.

\subsection{The conformal approach with the isotropic conformal Aristotelian algebra} \label{ssec:isoconfcalc}

As mentioned above, applying the conformal compensating technique with the isotropic conformal Aristotelian algebra entails constructing actions for two compensating scalar fields $\phi$, resp. $\psi$. We assume that their transformation rule under homogeneous conformal transformations only consists of a scaling under the dilatation $D_1$, resp. $D_2$, with weight $w_1$, resp. $w_2$: 
\begin{align}
\label{eq:phipsitrafo}
\delta \phi &= w_1 \sigma_1 \phi \,, \qquad \qquad \qquad \delta \psi = w_2 \sigma_2 \psi \,.
\end{align}
To build actions that are invariant under local homogeneous conformal transformations, we define the following covariant derivatives:
\begin{alignat}{2}
D_{A} \phi &\equiv \tau_{A}{}^{\mu} \left(\partial_\mu \phi - w_1 b_\mu \phi \right) \,, \qquad \qquad & D_{a} \phi &\equiv e_{a}{}^{\mu} \left(\partial_\mu \phi - w_1 b_\mu \phi \right) \,, \nonumber \\
D_{A} \psi &\equiv \tau_{A}{}^{\mu} \left(\partial_\mu \psi - w_2 c_\mu \psi \right) \,, \qquad \qquad & D_{a} \psi &\equiv e_{a}{}^{\mu} \left(\partial_\mu \psi - w_2 c_\mu \psi \right) \,.
\end{alignat}
Their transformation rules under homogeneous conformal transformations are given by:
\begin{align}
\delta D_{A} \phi &= \lambda_{A}{}^{B} D_{B} \phi + (w_1 - 1) \sigma_1 D_{A} \phi - 2 w_1 \lambda_{K A} \phi \,, \nonumber \\
\delta D_{a} \phi &= \lambda_{a}{}^{b} D_{b} \phi - \sigma_2 D_{a} \phi + w_1 \sigma_1 D_{a} \phi \,, \nonumber \\
\delta D_{A} \psi &= \lambda_{A}{}^{B} D_{B} \psi - \sigma_1 D_{A} \psi + w_2 \sigma_2 D_{A} \psi \,, \nonumber \\
\delta D_{a} \psi &= \lambda_{a}{}^{b} D_{b} \psi + (w_2 - 1) \sigma_2 D_{a} \psi - 2 w_2 \lambda_{K a} \psi \,.
\end{align}
We will also need the following definitions for the ``longitudinal Laplacian" $D^{A} D_{A} \phi$ and the ``transversal Laplacian" $D^a D_a \psi$:
\begin{align}
D^{A} D_{A} \phi &\equiv \tau_{A}{}^{\mu} \left(\partial_{\mu} D^{A} \phi - \omega_{\mu}{}^{AB} D_{B} \phi - (w_1 - 1) b_{\mu} D^{A} \phi + 2 w_1 f_{\mu}{}^{A} \phi \right) \,, \nonumber \\
D^{a} D_{a} \psi &\equiv e_{a}{}^{\mu} \left( \partial_{\mu} D^{a} \psi - \omega_{\mu}{}^{ab} D_{b} \psi - (w_2 - 1) c_{\mu} D^{a} \psi + 2 w_2 f_{\mu}{}^{a} \psi \right) \,.
\end{align}
From \eqref{eq:haconfrules0} and \eqref{eq:phipsitrafo}, one finds that these Laplacians transform as follows under homogeneous conformal transformations:
\begin{align} 
    \label{eq:trafoboxes}
\delta D^{A} D_{A} \phi &= (w_1 - 2) \sigma_1 D^{A} D_{A} \phi - 2 (2 w_1 + p - 1) \lambda_K{}^{A} D_{A} \phi \,, \nonumber \\
\delta D^{a} D_{a} \psi &= (w_2 - 2) \sigma_2 D^{a} D_{a} \psi - 2 (2 w_2 + D - p - 3) \lambda_K{}^{a} D_{a} \psi \,.
\end{align}

With these covariant derivatives and Laplacians, we can build three types of actions for $\phi$ and $\psi$ that are invariant under homogeneous conformal transformations. Upon gauge-fixing the dilatation and special conformal symmetries, these reduce to Aristotelian gravity theories of the electric, magnetic and electric-magnetic types. We will now discuss these three cases in turn.

\vspace{5pt}

\noindent\textbf{Electric Aristotelian gravity:}

\noindent As our first application of the conformal compensating technique for the algebra \eqref{eq:galg}, we consider the following action for the two compensating scalars $\phi$ and $\psi$, with covariant derivatives and transformation rules given in eqs. \eqref{eq:phipsitrafo}--\eqref{eq:trafoboxes}:
\begin{align}
    \label{eq:Lconfelel}
    \mathcal S_{1} &= \int d^D x\, \Omega \left[4 \alpha \psi^x  D^{a} \phi D_{a} \phi + 4 \beta \phi^y D^{A} \psi D_{A} \psi \right] \,,
\end{align}
where $\alpha,\beta\ne 0$  are real numbers. Here and in the following, we use the notation 
\begin{align}
\Omega={\rm det}(\tau_\mu{}^A,e_\mu{}^a) \,.
\end{align}
The exponents $x$ and $y$ are real numbers that are determined by requiring that \eqref{eq:Lconfelel} is invariant under the longitudinal and transversal dilatations $D_1$ and $D_2$. One can check that this is the case, provided that the weights $w_1$ and $w_2$ and the exponents $x$ and $y$ are given by:
\begin{align} \label{eq:Lconfelelws}
w_1 &= - \frac{(p+1)}{2} \,, \qquad w_2 = - \frac{(D-p-1)}{2} \,, \qquad 
x = 2 \frac{(D-p-3)}{(D-p-1)} \,, \qquad y = 2 \frac{(p-1)}{(p+1)} \,.
\end{align}
The action $S_1$ is moreover invariant under all homogeneous transformations of the algebra \eqref{eq:galg}, since it does not contain any fields that transform non-trivially under the special conformal transformations $K_A$ and $K_a$ and is manifestly invariant under $M_{AB}$ and $J_{ab}$. It is gauge-equivalent to an Aristotelian gravity action of the electric type, i.e., consisting of terms that are quadratic in intrinsic torsion tensor components. This is most easily seen by adopting the following gauge conditions to fix the two dilatations $D_1$ and $D_2$:
\begin{align} \label{eq:phipsiis1}
D_1 \ \text{gauge: } \ \ \phi = 1 \,, \qquad \qquad \qquad D_2 \ \text{gauge: } \ \ \psi = 1 \,.
\end{align}
Using this in \eqref{eq:Lconfelel}, along with \eqref{eq:confdepombc} and \eqref{eq:Lconfelelws}, one then obtains the following electric Aristotelian gravity action
\begin{equation}\label{eq:elecricpbraneAristotelian1}
\mathcal S^{(1)}_{\rm electric-AG} = \int d^D x\, \Omega \Big[ \alpha \, T_a{}^A{}_A T^{aB}{}_B   + \beta\, E_A{}^a{}_a E^{Ab}{}_b \Big] \,.
\end{equation}

This is not the only possible electric Aristotelian gravity action. Electric theories can be constructed out of squares of any of the intrinsic torsion components \eqref{intrinsic1}, as long as no hidden Galilei or Carroll boost invariance is introduced in the process. There exists in fact a larger number of electric Aristotelian theories than in the Carrollian or Galilean case, where the requirement of boost invariance restricts the number of possibilities for electric type actions, see \cite{Bergshoeff:2023rkk}. As an example of such an Aristotelian electric theory that is different from \eqref{eq:elecricpbraneAristotelian1}, we mention
\begin{equation}\label{eq:elecricpbraneAristotelian2}
\mathcal S^{(2)}_{\rm electric-AG}=  \int d^D x\, \Omega \, \left[\alpha\, T_{ab}{}{}^A T^{ab}{}{}_A + \beta \,E_{AB}{}{}^a E^{AB}{}{}_a\right] \,,
\end{equation}
where $\alpha,\beta\neq0$ and $1\leq \, p \, \leq D-3$. The first term of \eqref{eq:elecricpbraneAristotelian2} is invariant under Galilei boost transformations,\,\footnote{Denoting the parameters of $p$-brane Galilean, resp. Carrollian boosts by $\lambda_G{}^{Aa}$, resp. $\lambda_C{}^{Aa}$, one has that Galilean boosts act as $\delta e_\mu{}^a=-\lambda_{G A}{}^a \tau_\mu{}^A$ and $\delta \tau_\mu{}^A=0$, whereas Carrollian boosts act as $\delta \tau_\mu{}^A=\lambda_C{}^A{}_a e_\mu{}^a$ and $\delta e_\mu{}^a=0$.} but not under Carroll boosts. Likewise, the second term is invariant under Carroll but not under Galilei boosts. The total action \eqref{eq:elecricpbraneAristotelian2} thus does not exhibit any boost symmetry. Note that the first term in \eqref{eq:elecricpbraneAristotelian2} is the electric $p$-brane Galilei gravity action\,\footnote{We use the terms `electric', resp. `magnetic' for $p$-brane Galilei gravity to denote terms involving squares of intrinsic torsion tensors, resp. a curvature, in the same way as they are used for $p$-brane Carroll gravity.} found in \cite{Bergshoeff:2023rkk}, whereas the second term is the electric $p$-brane Carroll gravity action. The latter follows from its $(D-p-2)$-brane Galilean counterpart through the duality map
\begin{equation}\label{eq:dualitymap}
p \Longleftrightarrow D-p-2\,, \qquad  \tau_\mu{}^A \Longleftrightarrow
e_\mu{}^a\,, \qquad  \omega_{\mu}{}^A{}_B  \Longleftrightarrow \omega_{\mu}{}^a{}_b 
\,, \qquad \alpha \Longleftrightarrow \beta \,.
\end{equation}
This duality map also relates the first and the second terms of the action \eqref{eq:elecricpbraneAristotelian1}.

The intrinsic torsion components $T_{aA}{}^A$ and $E_{Aa}{}^{a}$ are the only ones whose transformations under the dilatations $D_1$ and $D_2$ contain non-covariant terms, i.e., terms that involve the derivatives of the dilatation parameters. All other intrinsic torsion components listed in \eqref{intrinsic1} undergo a simple rescaling with the dilatation parameters and are thus dilatation-covariant. Terms that only involve such dilatation-covariant intrinsic torsion components can be made dilatation invariant without the need for derivatives of the compensating scalars, by simply multiplying them with suitable powers of the compensating scalars. As a result, one finds that the action \eqref{eq:elecricpbraneAristotelian2} can not be written in a gauge-equivalent way as a dilatation invariant action for \emph{dynamical} scalars (i.e., involving terms that are at least quadratic in derivatives of the scalars), as in \eqref{eq:Lconfelel}. Instead, the action that is invariant under homogeneous conformal transformations and leads to \eqref{eq:elecricpbraneAristotelian2} upon using the gauge-fixing \eqref{eq:phipsiis1}, is given by:
\begin{align}\label{needed}
\int d^D x\, \Omega \left[\alpha \phi^{-p-3} \psi^{p+5-D} T_{ab}{}^{A} T^{ab}{}_{A} + \beta \phi^{3-p} \psi^{p-1-D} E_{AB}{}^{a} E^{AB}{}_{a} \right] \,.
\end{align}
Note that the powers of the compensating scalars $\phi$ and $\psi$ are really needed to achieve dilatation invariance: the action \eqref{eq:elecricpbraneAristotelian2} is by itself not invariant under any of the two dilatations of \eqref{eq:haconfrules0}.

A notable feature of electric Aristotelian gravity is that the coupling of the two compensating scalars only contains the dependent dilatation gauge field components $b_a$ and $c_A$ and no special conformal transformations are involved. This is characteristic for a result that can also be obtained from the conformal program based upon the smaller an-isotropic conformal Aristotelian algebra. In fact, at each stage one can make the following truncations in the above actions:
\begin{equation}
\phi = \psi\hskip 1truecm {\rm and}\hskip 1truecm b_A=c_a=0\,.
\end{equation}
This brings us back to the conformal program based upon the smaller algebra that will be discussed in the next subsection. We will then see that the action \eqref{needed} can be made invariant under the single dilatation of the smaller algebra, without making use of the single compensating scalar,  for $D=2$. This action in that case therefore describes 2D conformal electric Aristotelian gravity like 4D conformal gravity in the relativistic case. Note that the actions that we will construct below in the magnetic and electric-magnetic case contain the independent dilatation gauge field components
$b_A$ and $c_a$ that do transform under the special conformal transformations and therefore cannot be truncated.
\vspace{5pt}

\noindent\textbf{Magnetic Aristotelian gravity:}

\noindent As a second application of the conformal technique based on the algebra \eqref{eq:galg}, let us assume that $p$ does not take on any of the values $0$, $1$, $D-3$ and $D-2$ and consider the following action for the two scalar fields $\phi$ and $\psi$:
\begin{align}
    \label{eq:Lconfmagnmagn}
    \mathcal S_2 = \int d^D x \, \Omega \left[\alpha \psi^x \phi D^A D_A \phi + \beta \phi^y \psi D^a D_a \psi \right] \,,
\end{align}
with real numbers $\alpha,\beta\ne0$ and $x$, $y \in \mathbb{R}$. This action is manifestly invariant under $M_{AB}$ and $J_{ab}$, while invariance under the special conformal transformations $K_{A}$ and $K_{a}$ holds when the weights $w_1$ and $w_2$ are given by:
\begin{align}
    \label{eq:w1w2magnmagn}
w_1 &= - \frac{(p-1)}{2} \,, \qquad \qquad \qquad w_2 = - \frac{(D-p-3)}{2} \,.
\end{align}
If furthermore the exponents $x$ and $y$ are given by 
\begin{align} \label{eq:xymagnmagn}
    x &= 2 \frac{(D-p-1)}{(D-p-3)} \,, \qquad \qquad \qquad y = 2 \frac{(p+1)}{(p-1)} \,,
\end{align}
the action \eqref{eq:Lconfmagnmagn} is also invariant under the two dilatations $D_1$, $D_2$ and thus under all homogeneous conformal transformations. Fixing the special conformal transformations $K_{A}$, $K_{a}$ and the two dilatations $D_1$, $D_2$, using the gauge conditions
\begin{alignat}{2}
    D_1 \ \text{gauge: }& \ \ \phi = 1 \,, \qquad \qquad \qquad &  D_2 \ \text{gauge: }& \ \ \psi = 1 \,, \nonumber \\  K_{A} \ \text{gauge: }& \ \ b_{A} = 0\,, \qquad \qquad \qquad & K_{a} \ \text{gauge: }& \ \ c_{a} = 0 \,,
\end{alignat}
then shows that the action  \eqref{eq:Lconfmagnmagn} is gauge-equivalent to the following magnetic Aristotelian gravity action:
\begin{align} \label{eq:fixedmagnaction}
\int d^D x \, \Omega \left[\frac{\alpha}{4} \frac{(p-1)}{p} R_{AB}{}^{AB}(M) + \frac{\beta}{4} \frac{(D-p-3)}{(D-p-2)} R_{ab}{}^{ab}(J) \right] \,.
\end{align}
We will refer to the first, resp. second term as the ``longitudinal'', resp. ``transversal''  magnetic term. Note that the transversal magnetic term is reminiscent of the $p$-brane magnetic Galilei gravity action \cite{Bergshoeff:2023rkk}, while the longitudinal term resembles the $p$-brane magnetic Carroll gravity action (obtained from its $(D-p-2)$-brane Galilei counterpart via the duality map \eqref{eq:dualitymap}). It is however important to note that this is a mere resemblance and that the two terms of \eqref{eq:fixedmagnaction} are not equal to the $p$-brane magnetic Carroll and Galilei actions. In particular, the latter contain independent spin-connection components that act as Lagrange multipliers for intrinsic torsion constraints. In Aristotelian geometry, all spin-connection components are dependent and hence \eqref{eq:fixedmagnaction} contains no such Lagrange multiplier terms that constrain the intrinsic torsion.

Since one can not write down a longitudinal, resp. transversal magnetic term for $p=0$, resp. $p=D-2$, we have excluded these cases in the above discussion. We have also left out the values $p=1$ and $p=D-3$, because our conformal technique fails for them. This can be seen from the fact that the values of either $y$ or $x$ are not defined for $p=1$ or $p=D-3$ (see \eqref{eq:xymagnmagn}). Moreover, for $p=1$, the first term in \eqref{eq:fixedmagnaction} vanishes, while for $p = D-3$, the second term is not present. Note that in these cases, one of the two factors of the algebra \eqref{eq:galg} corresponds to a relativistic or Euclidean two-dimensional conformal algebra. The failure of the conformal approach for $p=1$ or $p=D-3$ is thus similar to what happens for the standard relativistic conformal technique in $D=2$ (where the Einstein-Hilbert term that would be constructed by applying this technique is a total derivative).

Even though the conformal approach does not work for $p=1$ and $p=D-3$, the corresponding longitudinal and transversal magnetic terms $\Omega R_{AB}{}^{AB}(M)$ and $\Omega R_{ab}{}^{ab}(J)$ exist and do not correspond to total derivatives. One can thus generalize the action \eqref{eq:fixedmagnaction} for all values of $p$ with $1 \leq p \leq D-3$:
\begin{equation}
\label{eq:magneticAristotelian}
\mathcal S_{\rm magnetic-AG} =  \int d^D x \, \Omega \left(\alpha R_{AB}{}^{AB}(M) + \beta R_{ab}{}^{ab}(J) \right) \,.
\end{equation}
This is then the most general action of magnetic Aristotelian gravity. Note that the transversal, resp. longitudinal magnetic term of $p$-brane Aristotelian geometry that appears in this action, can be obtained by applying the duality map \eqref{eq:dualitymap} to the longitudinal, resp. transversal magnetic term of $(D-p-2)$-brane Aristotelian geometry.

\vspace{5pt}

\noindent\textbf{Electric-Magnetic Aristotelian gravity:}

\noindent Finally, it is also possible to construct Aristotelian gravity actions of the mixed electric-magnetic type, that contain both a term quadratic in intrinsic torsion components and a curvature term. As an example of how such theories can be obtained from the conformal approach, we consider the following action:
\begin{align}
    \label{eq:Lconfelmagn}
    \mathcal S_3 &= \int d^D x \, \Omega \left[\alpha \phi^x \psi^y D^A \psi D_A \psi + \beta \phi^u \psi^w D^a D_a \psi \right] \,,
\end{align}
where $\alpha,\beta\ne 0$ are real numbers and $x$, $y$, $u$, $w \in \mathbb{R}$. This action is manifestly invariant under longitudinal Lorentz transformations $M_{AB}$ and transversal rotations $J_{ab}$ and trivially under longitudinal special conformal transformations $K_A$ (since it does not contain any fields that transform under $K_A$). From \eqref{eq:trafoboxes}, one sees that $S_3$ is invariant under transversal special conformal transformations $K_a$, provided the weight $w_2$ is given by:
\begin{align} 
    \label{eq:w2elmagn}
    w_2 = - \frac{(D-p-3)}{2} \,.
\end{align}
In order to ensure invariance under the dilatations $D_1$ and $D_2$, the weights $w_1$, $w_2$ and exponents $x$, $y$, $u$ and $w$ have to satisfy the following equations:
\begin{alignat}{2}
    \label{eq:dilcondelmagn}
    & x w_1 + p - 1 = 0 \,, \qquad \qquad \qquad & 
    & u w_1 + p + 1 = 0 \,, \nonumber \\
    & (y+2) w_2 + D - p - 1 = 0 \,, \qquad \qquad \qquad &
    & (w + 1) w_2 + D - p - 3 = 0 \,.
\end{alignat}
Together with \eqref{eq:w2elmagn}, the last two of these equations determine $y$ and $w$ as
\begin{align}
    \label{eq:ywelmagn}
    y = \frac{4}{(D-p-3)}\,, \qquad \qquad \qquad w = 1 \,.
\end{align}
The first two equations of \eqref{eq:dilcondelmagn} can be used to solve for $x$ and $u$ in terms of $w_1$ as follows:
\begin{align}
    \label{eq:xyelmagn}
    x = - \frac{(p-1)}{w_1} \,, \qquad \qquad \qquad u = - \frac{(p+1)}{w_1} \,.
\end{align}
The weight $w_1$ is then not fixed by any of the conditions \eqref{eq:w2elmagn}, \eqref{eq:dilcondelmagn} and can be arbitrarily chosen. This corresponds to a freedom to perform a field redefinition of the compensating scalar $\phi$, such that the redefined scalar has a different dilatation weight. 

With the values \eqref{eq:w2elmagn}, \eqref{eq:ywelmagn} and \eqref{eq:xyelmagn} for the weight $w_2$ and exponents $x$, $y$, $u$ and $w$, the action \eqref{eq:Lconfelmagn} is invariant under homogeneous conformal transformations. Fixing the $D_1$, $D_2$ and $K_{a}$ transformations, by adopting the gauge conditions:
\begin{align}
    D_1 \ \text{gauge: } \ \ \phi = 1 \,, \qquad \qquad D_2 \ \text{gauge: } \ \ \psi = 1 \,, \qquad \qquad K_{a} \ \text{gauge: } \ \ c_{a} = 0 \,,
\end{align}
one finds that \eqref{eq:Lconfelmagn} reduces to the following Aristotelian gravity action of electric-magnetic type:
\begin{align} \label{eq:Lfixelmagn1}
    \int d^D x \, \Omega \left[\frac{\alpha}{4} \frac{(D-p-3)^2}{(D-p-1)^2} E^A{}_{a}{}^{a} E_{Ab}{}^{b} + \frac{\beta}{4} \frac{(D-p-3)}{(D-p-2)} R_{ab}{}^{ab}(J)  \right] \,.
\end{align}

In a similar vein, one can consider the action
\begin{align}
    \label{eq:Lconfelmagn2}
    \mathcal S_4 &= \int d^D x \, \Omega \left[\beta \phi^x \psi^y D^a \phi D_a \phi + \alpha \phi^u \psi^w D^A D_A \phi \right] \,.
\end{align}
Proceeding as above, one finds that $S_4$ is invariant under homogeneous conformal transformations, provided the weight $w_1$ and exponents $x$, $y$, $u$ and $w$ are given by
\begin{align} \label{eq:xblowupelmagn2}
    x &= \frac{4}{(p-1)} \,, \qquad y = - \frac{(D-p-3)}{w_2} \,, \qquad u = 1 \,, \qquad w = - \frac{(D-p-1)}{w_2} \,, \nonumber \\
    w_1 &= - \frac{(p-1)}{2} \,,
\end{align}
where $w_2$ can be arbitrarily chosen. The action \eqref{eq:Lconfelmagn2} is gauge-equivalent to the following electric-magnetic Aristotelian gravity action
\begin{align} \label{eq:Lfixelmagn2}
    \int d^D x \, \Omega \left[\frac{\beta}{4} \frac{(p-1)^2}{(p+1)^2} T^a{}_{A}{}^{A} T_{aB}{}^{B} + \frac{\alpha}{4} \frac{(p-1)}{p} R_{AB}{}^{AB}(M)  \right] \,,
\end{align}
as is seen by fixing the dilatations $D_1$, $D_2$ and special conformal transformations $K_A$ with the gauge conditions:
\begin{align}
    D_1 \ \text{gauge: } \ \ \phi = 1 \,, \qquad \qquad D_2 \ \text{gauge: } \ \ \psi = 1 \,, \qquad \qquad K_{A} \ \text{gauge: } \ \ b_{A} = 0 \,.
\end{align}
Note that the $K_{a}$ transformations do not need to be fixed, since the action \eqref{eq:Lconfelmagn2} does not contain any fields that transform non-trivially under $K_{a}$. Note as well that the actions \eqref{eq:Lfixelmagn1} and \eqref{eq:Lfixelmagn2} are related to each other via the duality map \eqref{eq:dualitymap}.

The actions \eqref{eq:Lfixelmagn1} and \eqref{eq:Lfixelmagn2} are not the most general Aristotelian gravity theories of the mixed electric-magnetic type. They can be generalized in two ways. First, as in the magnetic case, the above conformal construction fails for $p=1$ or $p=D-3$, when one of the factors of the algebra \eqref{eq:galg} becomes a two-dimensional relativistic or Euclidean conformal algebra. In particular, for $p=1$ (resp. $p=D-3$), the exponent $x$ in \eqref{eq:xblowupelmagn2} (resp. $y$ in \eqref{eq:ywelmagn}) is not defined and the gauge-fixed action \eqref{eq:Lfixelmagn2} (resp. \eqref{eq:Lfixelmagn1}) vanishes. As in the magnetic case, this failure is an artefact of the conformal approach and one can generalize the actions \eqref{eq:Lfixelmagn1} and \eqref{eq:Lfixelmagn2} to
\begin{align} \label{eq:genelmagngrav1}
 \int d^D x \, \Omega \left[\alpha E^A{}_{a}{}^{a} E_{Ab}{}^{b} + \beta R_{ab}{}^{ab}(J)  \right] \quad \text{and} \quad
 \int d^D x \, \Omega \left[\beta T^a{}_{A}{}^{A} T_{aB}{}^{B} + \alpha R_{AB}{}^{AB}(M)  \right] \,,
\end{align}
respectively, that are valid electric-magnetic Aristotelian gravity actions for all values of $p$ between 1 and $D-3$ and that are related to each other via the duality map \eqref{eq:dualitymap}. 

The actions \eqref{eq:genelmagngrav1} only involve the intrinsic torsion tensor components that do not transform covariantly under the dilatations $D_1$ and $D_2$. Like in the purely electric case, this is not the only possibility and one is free to generalize \eqref{eq:genelmagngrav1} by considering squares of any of the intrinsic torsion components \eqref{intrinsic1}, as long as they do not introduce a hidden Galilei or Carroll boost invariance. For example, instead of \eqref{eq:genelmagngrav1}, the following actions
\begin{align} \label{eq:genelmagngrav2}
 \int d^D x \, \Omega \left[\alpha E_{AB}{}^{a} E^{AB}{}_{a} + \beta R_{ab}{}^{ab}(J)  \right] \quad \text{and} \quad
 \int d^D x \, \Omega \left[\beta T_{ab}{}^{A} T^{ab}{}_{A} + \alpha R_{AB}{}^{AB}(M)  \right] \,,
\end{align}
that involve squares of the dilatation-covariant intrinsic torsion tensor components $T_{ab}{}^{A}$ and $E_{AB}{}^{a}$, can also be considered as electric-magnetic Aristotelian gravity actions. They are valid for all values of $p$ with  $1 \leq p \leq D-3$ and are related via the duality map \eqref{eq:dualitymap}. As in the electric case, the terms $\Omega E_{AB}{}^{a} E^{AB}{}_a$ and $\Omega T_{ab}{}^{A} T^{ab}{}_{A}$ can be made invariant under the dilatations $D_1$ and $D_2$ by multiplying them with appropriate powers of $\phi$ and $\psi$. They are thus not obtained in the conformal approach by gauge-fixing terms that involve derivatives of the compensating scalars $\phi$ and $\psi$.

\vspace{8pt}

\subsection{The an-isotropic conformal Aristotelian algebra} \label{ssec:anisoconfcalc}

Instead of applying the conformal compensating technique with the isotropic conformal Aristotelian algebra \eqref{eq:galg}, it is also possible to use the an-isotropic algebra \eqref{eq:galganiso}. As mentioned in the introduction of this section, only electric Aristotelian gravity theories can be obtained in this way, since the an-isotropic algebra does not contain any special conformal transformations. Moreover, we only need one real scalar, since the algebra \eqref{eq:galganiso} only contains one dilatation that needs to be fixed. We will denote this scalar by $\rho$.

We assume that under the homogeneous conformal transformations of \eqref{eq:galganiso}, $\rho$ only scales under the dilatation $D$ with weight $w$: 
\begin{equation}
\delta \rho = w\lambda_D \rho\,.
\end{equation}
To build conformally invariant actions for $\rho$, we need the following covariant derivatives:
\begin{equation}\label{eq:CovD}
D_A \rho =\, \tau_A{}^\mu \partial_\mu \rho- w \,b_A \rho\,, \qquad D_a \rho=\, e_a{}^\mu \partial_\mu \rho -w \,b_a \rho \,,
\end{equation}
where $b_A=\tau_A{}^\mu b_\mu$, $b_a=e_a{}^\mu b_\mu$ are the dependent dilatation gauge fields given in \eqref{eq:Conformalb1} and \eqref{eq:Conformalb2}. These derivatives transform as follows under the homogeneous conformal transformations of \eqref{eq:galganiso}:
\begin{align}
\delta D_A \rho &= \lambda_{A}{}^{B} D_{B} \rho + (w - z) \lambda_D D_{A} \rho \,, \nonumber \\
\delta D_a \rho &= \lambda_{a}{}^{b} D_{b} \rho + (w - 1) \lambda_D D_{a} \rho \,.
\end{align}
We can then consider the following action for the scalar $\rho$:
\begin{equation}
\label{eq:RealScalarAc}
\mathcal S_\rho  = \frac12 \int d^Dx\, \Omega\Big(c_1\,D_A\rho D^A\rho + c_2\,D_a\rho D^a\rho\Big) \,,
\end{equation}
where the real numbers $c_1$ and $c_2$ are both non-zero and $c_1 \neq c_2$. This action is invariant under the homogeneous conformal transformations of \eqref{eq:galganiso} provided that $z$ and $w$ are given by 
\begin{equation}\label{zandw-quadcomplexscalar}
z=1\,,\qquad w=\, \frac{1}{2}(2-D) \,.
\end{equation}
Indeed, invariance of \eqref{eq:RealScalarAc} under longitudinal Lorentz transformations $M_{AB}$ and transversal rotations $J_{ab}$ is manifest, while \eqref{zandw-quadcomplexscalar} ensures invariance under the dilatation $D$.

The action \eqref{eq:RealScalarAc} is gauge-equivalent to the electric Aristotelian gravity action \eqref{eq:elecricpbraneAristotelian1}, as is seen by imposing the gauge condition  
\begin{equation}\label{eq:gaugefixingphi}
\rho =\ 1 \,,
\end{equation}
to fix the local dilatation $D$. 
Upon substituting this condition into the action \eqref{eq:RealScalarAc}, we obtain the electric Aristotelian action \eqref{eq:elecricpbraneAristotelian1} introduced in the previous subsection:
\begin{equation}\label{eq:elecricpbraneAristotelianV2}
\begin{aligned}
\mathcal S_{\rm electric-AG} &= \frac12\int d^D x\, \Omega \Big(c_1\, w^2 \, b_A b^A  + c_2\, w^2 \, b_a b^a \Big) 
\\&=\int d^D x\, \Omega \Big( \alpha\, T_a{}^A{}_A T^{aB}{}_B + \beta \, E_A{}^a{}_a E^{Ab}{}_b \Big) \,,
\end{aligned}
\end{equation}
with
\begin{equation}
\beta =\, \frac{c_1 (D-2)^2}{8(D-p-1)^2} \,, \qquad \alpha=\,  \frac{c_2(D-2)^2}{8(p+1)^2} \,.
\end{equation}

Terms in electric Aristotelian gravity theories that involve squares of the dilatation-covariant intrinsic torsion components $T_{a}{}^{\{A,B\}}$, $T_{ab}{}^{A}$, $E_{A}{}^{\{a,b\}}$ and $E_{AB}{}^{a}$ can be made dilatation-invariant by multiplying them with appropriate powers of the compensating scalar $\rho$. For instance, the action \eqref{eq:elecricpbraneAristotelian2} can, for $D\neq2$ and for the values \eqref{zandw-quadcomplexscalar} for $z$ and $w$, be made invariant under the homogeneous conformal transformations of \eqref{eq:galganiso} as follows\,\footnote{One can also relax the requirement that $z$ and $w$ are given by \eqref{zandw-quadcomplexscalar}. Considering an action of the form
\begin{align}
\int d^D x\, \Omega \left[\alpha \rho^x T_{ab}{}^{A} T^{ab}{}_{A} + \beta \rho^y E_{AB}{}^{a} E^{AB}{}_{a} \right] \,,
\end{align}
with $x$, $y \in \mathbb{R}$, one finds that it is invariant under the homogeneous conformal transformations of \eqref{eq:galganiso}, provided that $z(p+1) + D - p - 5 + x w + 2 z = 0$ and $(x-y) w + 6 (z - 1) = 0$.}
\begin{align}
    \label{eq:anisoelgrav2}
\int d^D x\, \Omega \left[\alpha \rho^2 T_{ab}{}^{A} T^{ab}{}_{A} + \beta \rho^2 E_{AB}{}^{a} E^{AB}{}_{a} \right] \,.
\end{align}
In this way, the action \eqref{eq:elecricpbraneAristotelian2} is retrieved from the conformal approach based on the an-isotropic algebra \eqref{eq:galganiso}, by adopting the gauge condition \eqref{eq:gaugefixingphi} in \eqref{eq:anisoelgrav2}. As we already mentioned when discussing the electric case in the previous subsection, when $D=2$ (and $z=1$) there is no need for the insertion of suitable powers of the compensating scalar $\rho$, since in that case \eqref{eq:elecricpbraneAristotelian2} is already invariant under the dilatation $D$ of \eqref{eq:galganiso}. This case can thus be viewed as an electric Aristotelian analogue of four-dimensional conformal gravity.

\section{Matter couplings} \label{sec:confargrav}

In this section, we apply the conformal approach to construct examples of Aristotelian gravity with matter. We discuss several models that are quadratic in derivatives as well as higher-derivative models, inspired by similar structures studied in fracton theories.

For the quadratic-derivative case, we have already seen in the previous section how gauging the conformal Aristotelian symmetry for a real scalar field leads to an electric Aristotelian action. Similarly, we will see in this section that the gauging of the conformal Aristotelian symmetry for a higher-derivative scalar field  will lead to an 
 electric higher-derivative Aristotelian gravity action.

In both the quadratic and higher-derivative case, we will add a second scalar field that compensates for a U(1) symmetry in the quadratic-derivative case and for a U(1) and dipole symmetry in the higher-derivative case. Introducing gauge fields for these symmetries we will consider in the quadratic-derivative case an Aristotelian version of electrodynamics based upon a single vector field that can be straightforwardly coupled to Aristotelian gravity.
By contrast, in the higher-derivative case we will introduce a vector gauge field with longitudinal components and a symmetric tensor gauge field with transverse components. The corresponding kinetic terms for these gauge fields can only be coupled to Aristotelian gravity by making use of the scalar that compensates for the dipole gauge symmetries.

In the models considered below we will combine the compensating scalar for the gravitational dilatations and the compensating scalar for the U(1) and dipole gauge symmetries into a complex scalar field.
We will consider the gauged versions of these complex scalar field models for three different cases: (i) we first gauge the conformal Aristotelian symmetry using the Aristotelian gravitational gauge fields we introduced in the previous sections;  (ii) we separately gauge the $U(1)$ and dipole symmetries using the vector and symmetric tensor gauge fields mentioned above and, finally,  (iii) we combine (i) and (ii) and gauge all symmetries simultaneously.

\subsection{Quadratic-derivative models}
\label{quadracticmodels}

The quadratic-derivative complex scalar field model that we will discuss in this subsection not only requires knowledge of Aristotelian gravity that we already discussed in the previous section, but also makes use of the notion of Aristotelian electrodynamics that we will discuss separately below before continuing with discussing the quadratic-derivative complex scalar field model.
\vskip .2truecm
\noindent\textbf{Aristotelian electrodynamics:} 

\noindent We  consider the $p$-brane Aristotelian version of electrodynamics in $D$-dimensional flat spacetime, that we take to be given by the following general combination of terms that are quadratic in components of the field strength of a gauge potential $A_{\mu} = (A_A, A_a)$:
\begin{equation}\label{LAEflat}
\mathcal S_{AE}=-\frac12 \int d^D x \,
\left(\frac{b_0}2\,F_{AB}\,F^{AB} + b_1\,F_{Aa}\,F^{Aa} +\frac{b_2}{2}\,F_{ab}\, F^{ab}\right)\,,
\end{equation}
where $b_0,b_1,b_2$ are real numbers and  the different components of the field strength read
\begin{equation}
F_{AB}=\partial_A A_B-\partial_B  A_A, \qquad 
F_{Aa}=\partial_A A_a-\partial_a  A_A, \qquad F_{ab}=\partial_a A_{b}- \partial_b A_{a}\,.
\end{equation} 
We take $b_0$, $b_1$ and $b_2$ not all equal in order to avoid the relativistic case. Since the first and the third terms are respectively invariant under Carroll and Galilei boosts, the requirement of no boost invariance for $0 < p < D-2$ is either $b_1\neq 0$ or $b_0,b_2\neq 0$. For $p=0$ the term along $b_0$ vanishes while the term along $b_1$ becomes Carroll boost invariant, leading to the requirement $b_1,b_2\neq0$. Similarly, for $p=D-2$ we need to impose $b_0,b_1\neq0$ since the term along $b_2$ vanishes and the term along $b_1$ becomes Galilei boost invariant. 
The first and third terms in \eqref{LAEflat} are related by a duality map of the form \eqref{eq:dualitymap}, with $\alpha$ and $\beta$ replaced by $b_0$ and $b_2$, respectively, whereas the term proportional to $b_1$ maps into itself under this transformation. 

The transformation rules of the gauge fields under local $U(1)$ transformations and global rotations and dilatations are given by 
\begin{equation}\label{Atrafos}
\delta A_A= \partial_A \lambda_{U(1)}+\lambda_A{}^B A_B-z\lambda_D A_A, \qquad
\delta A_a=\partial_a \lambda_{U(1)}+\lambda_a{}^b A_b-\lambda_D A_a \,,
\end{equation}
where the transformation under rotations and dilatations is inherited from \eqref{eq:localtrafosTorsionsDilatation1}. Under dilatations, the curvatures transform as
\begin{equation}
\delta F_{AB} = -2z \lambda_D  F_{AB}\,,\qquad
\delta F_{Aa} = -(z+1) \lambda_D  F_{Aa}\,,\qquad
\delta F_{ab} = -2 \lambda_D F_{ab}  \,.
\end{equation}
Thus, the action \eqref{LAEflat} is dilatation invariant only for $z=1$, as in \eqref{zandw-quadcomplexscalar}, and for $D=4$. When coupling the system to matter currents $J^A$ and $J^a$, the variation of the full action, up to total derivatives, takes the form
\begin{equation}\label{varLAEflatwithJ}
\delta \mathcal S= \delta \mathcal S_{AE} +\int d^D x\,\left(\delta A_A J^A+ \delta A_a J^a\right) \,.
\end{equation}
The field equations take the form
\begin{equation}\label{equationsAJ}
\begin{aligned}
\delta A_A\,&: \qquad && b_0 \,\partial_B  F^{AB}+b_1 \,\partial_a  F^{Aa}
=J^A \,,
\\[5pt]
\delta A_a\,&: && b_2 \partial_b F^{ab}-b_1\,\partial_A F^{Aa}= J^a \,,
\end{aligned}
\end{equation}
which can be combined to find the continuity equation 
\begin{equation}
\partial_A J^A + \partial_a J^a=0\,.
\end{equation}
Notice that despite the lack of boost invariance, the continuity equation takes the same form as in the relativistic case and does not depend on the constants $b_0$, $b_1$ and $b_2$.

This model can be coupled to Aristotelian gravity in a straightforward way by defining the curved space extension of \eqref{LAEflat}:
\begin{equation}\label{LAEcurved}
\mathcal S^c_{AE}=-\frac12 \int d^D x \,\Omega
\left(\frac{b_0}{2}\, F_{AB}\,  F^{AB}+b_1\, F_{Aa}\, F^{Aa} +\frac{b_2}{2}\, F_{ab}\,  F^{ab}\right)\,,
\end{equation}
where the superscript $c$ stands for ``curved-space'', and we distinguish between flat and curved indices as follows:
\begin{equation} F_{AB}=\tau_A{}^\mu \tau_B{}^\nu F_{\mu\nu}, \qquad  F_{Aa}=\tau_A{}^\mu e_a{}^\nu F_{\mu\nu}
, \qquad  F_{ab}=e_a{}^\mu e_b{}^\nu F_{\mu\nu},\qquad F_{\mu\nu}=\partial_\mu A_\nu -\partial_\nu A_\mu \,.
\end{equation}
As before, the system is $U(1)$-invariant by construction, whereas there is local dilatation invariance only when $D=4$ and $z=1$. In this case we can consider the Aristotelian gravity-Maxwell system
\begin{equation}\label{AristGravMax}
\mathcal S_{AGM}=  \mathcal S_{\rm electric-AG}+\mathcal S^c_{AE}\,,
\end{equation}
with the electric Aristotelian gravity action given in eq.~\eqref{eq:elecricpbraneAristotelianV2}. As we will see in the following, a Proca-like extension of this action can be obtained from a complex scalar field coupled to Aristotelian electrodynamics on a conformal Aristotelian geometry upon gauge-fixation.

\vspace{8pt}
\noindent\textbf{Complex scalar field:}

\noindent We are now ready to discuss the quadratic-derivative complex scalar field model. The real scalar field model given in eq.~\eqref{eq:RealScalarAc} can be generalized to the case of a complex scalar field $\Phi$ described by the action
\begin{equation}\label{eq:Lcomplexscalar}
\mathcal S_\Phi= \frac12 \int d^D x\left( c_1\, \partial_A \Phi \partial^A \Phi^* +c_2\, 
\partial_a \Phi \partial^a \Phi^*\right)\,
\end{equation}
which is invariant under global $U(1)$ transformations and dilatations with parameters $\lambda_{U(1)}$ and $\lambda_D$, respectively 
\begin{equation}\label{deltaPhiU1andD}
\delta \Phi = \left(i \lambda_{U(1)}  + w\lambda_D \right) \Phi\,.
\end{equation}
As before, the global dilatation invariance holds provided $w=(D-2)/2$ and $z=1$.

It is convenient to express the complex scalar field in terms of new variables $\rho$ and $\theta$ corresponding to the radius and the phase of $\Phi$, i.e.
\begin{equation}\label{phirhotheta}
\Phi = \rho\,e^{i\theta}\,.
\end{equation}
The action \eqref{eq:Lcomplexscalar} can then be written as
\begin{equation}
\mathcal S_\Phi = \frac12 \int d^D x \,
\bigg(c_1\, \partial_A \rho \partial^A\rho + c_2\, \partial_a \rho \partial^a\rho + \rho^2\left(c_1\, \partial_A\theta \partial^A\theta + c_2 \,\partial_a\theta \partial^a\theta\right)\bigg)
\end{equation}
and the transformations \eqref{deltaPhiU1andD} take the form of
\begin{equation}
\delta\rho= w \lambda_D\rho, \qquad
\delta\theta= \lambda_{U(1)}\,.
\end{equation}
In the following, we consider the gauging of these global symmetries, which allows one to construct an action for Aristotelian gravity coupled to a massive extension of Aristotelian electrodynamics upon gauge fixing. As mentioned at the beginning of this section we will do this gauging in three stages.

\vspace{8pt}
\noindent (i) \textit{ Gauging the conformal Aristotelian symmetry}: Coupling the complex scalar field model \eqref{eq:Lcomplexscalar} to conformal Aristotelian gravity produces the curved space action
\begin{equation}\label{eq:gaugedLcomplexscalar}
\mathcal S^c_\Phi=\frac12\int d^D x\,\Omega\left(c_1\,D_A \Phi D^A \Phi^* + c_2\, D_a \Phi D^a \Phi^*\right)\,,
\end{equation}
where $D_A \Phi$ and $D_a \Phi$ are defined via \eqref{eq:CovD}. We gauge-fix the dilatations by imposing the following gauge condition
\begin{equation}\label{gaugefixingrho}
\rho=1\; \Rightarrow \; \Phi = e^{i\theta}\,,
\end{equation}
which leads to the following action for a real scalar field $\theta$ coupled to Aristotelian gravity
\begin{equation}
\mathcal S=\mathcal S_{\rm electric-AG}
+\mathcal S^c_\theta \,,
\end{equation}
with
\begin{equation} 
\mathcal S^c_\theta = \frac12 \int d^D x \,\Omega\left(c_1\,\tau^{A \mu} \tau_{A}{}^{\nu} \partial_{\mu} \theta \partial_{\nu} \theta + c_2\, e^{a \mu} e_{a}{}^{\nu} \partial_{\mu} \theta \partial_{\nu} \theta\right) \,.
\end{equation}
This is an example of a matter-coupled Aristotelian gravity system consisting of a real scalar $\theta$ coupled to electric Aristotelian gravity described by the action  \eqref{eq:elecricpbraneAristotelianV2}.

\vspace{8pt}

\noindent (ii) \textit{Gauging the $U(1)$ symmetry}: The complex scalar $\Phi$ described by the action \eqref{eq:Lcomplexscalar} can be naturally coupled to the set of two $U(1)$  gauge fields of Aristotelian electrodynamics, $A_A$ and $A_a$, by considering the action
\begin{equation}\label{ScalarU(1)gauged}
\begin{aligned}
\mathcal S=\frac12\int d^Dx\,& \bigg[c_1\left(\partial_A-iA_A\right) \Phi \left(\partial^A+iA^A\right)  \Phi^* + c_2\left(\partial_a-iA_a\right) \Phi \left(\partial^a+iA^a\right)  \Phi^* 
\\&-|\Phi \Phi^*|^{\gamma/2}\left(\frac{b_0}{2}\, F_{AB} F^{AB}+ b_1\,F_{Aa}\,F^{Aa} +\frac{b_2}{2}\,F_{ab}\, F^{ab}\right)\bigg]\,.
\end{aligned}
\end{equation}
Here, $\gamma$ is a constant that parametrizes a non-minimal coupling between the scalar $\Phi$ and the Aristotelian electrodynamics action \eqref{LAEflat}. This coupling is needed to maintain conformal Aristotelian invariance. Indeed, using  \eqref{zandw-quadcomplexscalar}, one finds that dilatation invariance fixes $\gamma$ as
\begin{equation}\label{gamma}
\gamma = \frac{4-D}{w} \,.
\end{equation}
Since $\gamma$ vanishes for $D=4$, in the four-dimensional case the dilaton-like coupling in \eqref{ScalarU(1)gauged} disappears and the scalar $\Phi$ couples minimally to Aristotelian electrodynamics \eqref{LAEflat}, while still preserving Aristotelian conformal invariance. This resembles the conformal invariance of Maxwell's theory in the relativistic case. In the particular case $D=4$, the variation of the action \eqref{ScalarU(1)gauged} takes the form \eqref{varLAEflatwithJ} and the Aristotelian Maxwell equations take the form \eqref{equationsAJ} with
\begin{equation}
\begin{aligned}
J^A&=\frac{i c_1}2 \bigg(\Phi^* \left(\partial^A-iA^A\right)\Phi-\Phi \left(\partial^A+iA^A\right)\Phi^* \bigg)\,,
\\
J^a&=\frac{i c_2}2 \bigg(\Phi^* \left(\partial^a-iA^a\right)\Phi-\Phi \left(\partial^a+iA^a\right)\Phi^* \bigg)\,.
\end{aligned}
\end{equation}
This model for Aristotelian scalar electrodynamics can be generalized to higher dimensions albeit at the cost of losing dilatation symmetry.

Going back to the general case and adopting the variables \eqref{phirhotheta}, imposing the gauge-fixing condition
\begin{equation}\label{gaugefixingtheta}
\theta=0\; \Rightarrow \; \Phi =\rho
\end{equation}
leads to the following action for the real Aristotelian scalar field $\rho$ non-minimally coupled to the Aristotelian electrodynamics gauge fields:
\begin{equation}
\begin{aligned}
\mathcal S= \frac12 \int & d^D x \,
\bigg[c_1\, \partial_A \rho \partial^A\rho + c_2\, \partial_a \rho \partial^a\rho 
\\&-\rho^\gamma \left(\frac{b_0}{2}\, F_{AB} F^{AB}+b_1\,F_{Aa}\,F^{Aa} +\frac{b_2}{2}\,F_{ab}\, F^{ab}\right) + \rho^2\left(c_1 A_A A^A  +c_2 A_aA^a\right)\bigg]\,
\end{aligned}
\end{equation}
with $\gamma$ given by \eqref{gamma}.
Even though the $U(1)$ gauge symmetry has been gauge-fixed, this action is still invariant under global dilatations. In the following, we will gauge these dilatations thereby coupling the gauge fields $A_A$ and $A_a$ to Aristotelian gravity.

\vspace{8pt}

\noindent (iii) \textit{Gauging all symmetries considered in cases (i) and (ii)}:  We  now gauge the conformal Aristotelian symmetry and consider  the following generalization of the action \eqref{eq:gaugedLcomplexscalar}
\begin{equation}\label{fullgaugedscomplexcalar}
\begin{aligned}
\mathcal S=\frac12\int d^D x\, &\Omega\bigg[c_1\left(D_A-iA_A\right) \Phi \left(D^A+iA^A\right)  \Phi^* 
+ c_2\left(D_a-iA_a\right) \Phi \left(D^a+iA^a\right)  \Phi^* 
\\&-|\Phi \Phi^*|^{\gamma/2}\left(\frac{b_0}{2}\,  F_{AB} F^{AB}+b_1\, F_{Aa}\, F^{Aa} +\frac{b_2}{2}\, F_{ab}\, F^{ab}\right)\bigg]\,,
\end{aligned}
\end{equation}
where $\gamma$ is given by eq.~\eqref{gamma}. The coupling in the last line between the curved-space Aristotelian electrodynamics action \eqref{LAEcurved} and $\Phi$ is determined by requiring local dilatation invariance.   Fixing the local U(1) and dilatation symmetry by imposing the gauge-fixing condition
\begin{equation}\label{gaugefixPhi}
\Phi =1\,,
\end{equation}
we find that the action \eqref{fullgaugedscomplexcalar} reduces to
\begin{equation}
\mathcal S=\mathcal S_{\rm electric-AG}+\mathcal S^c_{AE} +\frac12 \int d^D x\,\Omega\left(c_1 \,A_A A^A  +c_2 \, A_aA^a\right)\,.
\end{equation}
This action is an extension of \eqref{AristGravMax} where the $U(1)$ symmetry is broken due to the presence of the Proca-like mass terms.  It describes electric Aristotelian gravity, given by the action  \eqref{eq:elecricpbraneAristotelianV2}, coupled to massive Aristotelian electrodynamics.

\vspace{8pt}
%%%%%%%%%%%%%%%%%%%%%%%%%%%%%%%%%%%%%%%
\subsection{Higher-derivative models }
The lack of boost invariance allows to consider a generalization of the Aristotelian scalar field action \eqref{eq:RealScalarAc} to include different numbers of derivatives in the longitudinal and transverse directions, while preserving invariance under the symmetries corresponding to the homogeneous part of the an-isotropic conformal Aristotelian algebra \eqref{eq:galganiso}. Examples of this type of Aristotelian field theories have been considered in the context of fractons \cite{Pretko:2016kxt,Pretko:2018jbi}. We will generalize this type of models to the case of Aristotelian $p$-brane foliations. In contrast to the quadratic-derivative models, the gauging of the $U(1)$ symmetry leads in this case to Aristotelian gauge fields that include higher-rank tensors, while the gauging of the Aristotelian conformal symmetry results into a higher-derivative Aristotelian gravity action. Below, we will consider three models of increasing complexity. First we will give a higher-derivative model for a real scalar field and show that the conformal program for this model leads to a higher-derivative generalization of the electric Aristotelian gravity action. Next, we will discuss a generalization of the Aristotelian electrodynamics given in the previous subsection that is based upon a vector and a symmetric tensor gauge field and is called a scalar-charge theory in the fracton literature \cite{Pretko:2016kxt,Pretko:2018jbi}. Finally, we consider a $p$-brane generalization of a so-called fracton field theory, i.e.~a higher-derivative field theory of a complex scalar where we will gauge both the gravitational dilatations as well as the U(1) and dipole symmetries. As in the previous subsection, we will do this gauging in three stages. First we will gauge the conformal Aristotelian symmetries, next the U(1) and dipole symmetries and finally we will perform these two gaugings simultaneously. In the latter case, we will also couple the complex scalar to the scalar-charge
theory. 
\vspace{8pt}

\noindent\textbf{Real higher-derivative scalar model:}

\noindent We consider a model for a real scalar field $\rho$ in flat Aristotelian spacetime that is second order in the  longitudinal derivatives and fourth order in the transversal derivatives, reminiscent of the fracton model, and is described by the action
\begin{equation}\label{eq:LrhoHD}
\mathcal{S}^{\rm HD}_\rho =\, \frac{1}{2}\int d^{D}x \Big(  c_1\, \partial_A \rho \partial^A \rho + c_2\, X_{ab} X^{ab} \Big)  \,,
\end{equation}
where $c_1\,, c_2$ are real coupling constants and $X_{ab}$ is defined by
\begin{equation}
X_{ab} = \partial_a\rho \partial_b\rho - \rho \partial_a \partial_b\rho\,.
\end{equation}
This action is invariant under global dilatations provided that the dynamical critical exponent $z$ and scaling weight $w$ are fixed to be
\begin{equation}\label{eq:zandw}
z=\frac{p-3-D}{p-3},\qquad \qquad \qquad w=\frac{p-3+D}{p-3} \,.
\end{equation} 
As in the quadratic-derivative case, the coupling of the real scalar to conformal Aristotelian gravity is done by replacing the partial derivatives with the covariant ones given in eq.~\eqref{eq:CovD}. In order to derive an expression for the second order covariant derivative, one uses that the transverse covariant derivative of $\rho$ transforms as follows:
\begin{equation}\label{eq:DDphi}
\delta D_a \rho = \lambda_a{}^b D_b \rho + (w-1)\lambda_D D_a \rho\,.
\end{equation}
Using this transformation rule, one derives that  the double covariant derivative of the scalar field takes the form
\begin{equation}\label{eq:DDphi2}
D_a D_b \rho = \partial_a D_b \rho - \omega_{ab}{}^c(\tau, e, b) D_c \rho - (w-1)b_a D_b \rho \,, 
\end{equation}
where $\omega_{ab}{}^c(\tau, e, b)$ and $b_a$ are the dependent gauge fields given in 
eqs.~\eqref{solutionsconformal3} and \eqref{eq:Conformalb1}, respectively. The resulting curved space action reads
\begin{equation}
\mathcal{S}^{\rm HD}_\rho=\, \frac{1}{2}\int d^Dx\,\Omega \left( c_1\, D_A \rho D^A \rho + c_2\, X_{ab}  X^{ab}
\right) \,,
\end{equation}
where now $X_{ab}$ is defined as the symmetric tensor\footnote{Note that the generalization of $X_{ab}$ to the curved space case is ambiguous since the covariant derivatives no longer commute. Here, we resolve this ambiguity by defining $X_{ab}$ as symmetric, which is well suited for the generalization we will introduce in the next section.}
\begin{equation}
X_{ab} = D_a\rho D_b\rho - \rho D_{(a} D_{b)}\rho\,.
\end{equation}
By imposing the gauge condition \eqref{eq:gaugefixingphi}, the action reduces to the following higher-derivative Aristotelian (HDAG) electric gravity theory 
\begin{equation}\label{eq:Agravityhigher}
\mathcal S_{\rm HDAG}=\, \frac{1}{2}\int d^D x\, \Omega\, \left(c_1\, w^2\,b_A b^A 
+c_2\,X_{ab} X^{ab}
\right) \,,
\end{equation}
where $X_{ab}$ is now given by
\begin{equation}
X_{ab} = w\left(
\mathcal D_{(a} b_{b)} + 2 b_a b_b -\delta_{ab} b_c b^c\right) \,,
\end{equation}
and $\mathcal D_a$ is the covariant derivative constructed with the non-conformal spin-connection $\omega_{a,bc}(\tau, e)$ defined in eq.~\eqref{solutions3}.

\vspace{8pt}
\noindent\textbf{Scalar-charge theory:} 

\noindent Below, we will generalize the real higher-derivative scalar model to a higher-derivative model for a complex field and gauge the dilations, U(1) and dipole symmetries of this model. The gauge fields for the U(1) and dipole symmetries are given by a vector field in the longitudinal directions and a symmetric tensor gauge field in the transverse directions, respectively. Before performing these gaugings, we will first consider the kinetic terms for these gauge fields that will replace the kinetic terms of Aristotelian electrodynamics that we encountered in the quadratic-derivative case. This generalized electromagnetic action is called a scalar-charge theory in the fracton literature \cite{Pretko:2016kxt,Pretko:2018jbi}.  

The gauging of the U(1) and dipole symmetries that we will give below can be achieved by two gauge fields, $A_A$ and $A_{ab}$, that transform  under local gauge transformations as follows:
\begin{equation}\label{fractontrafos}
\delta A_A= \partial_A \lambda_{U(1)}-z\lambda_D A_A, \qquad
\delta A_{ab}=\partial_a \partial_b \lambda_{U(1)}-2\lambda_D A_{ab} \,,
\end{equation} 
where $\lambda_{U(1)}$ and $\lambda_D$ are the gauge parameters of U(1)/dipole symmetries and dilatations, respectively. An extension of the standard form of the electric and magnetic field for these gauge fields reads
\begin{equation}
\label{eq:ScalarChargeTheoryF}
F_{Aab}=\partial_A A_{ab}-\partial_a \partial_ b A_A, \qquad F_{ab,c}=\partial_a A_{bc}- \partial_b A_{ac}\,.
\end{equation}
These curvatures  can be used to construct the following gauge-invariant action in flat spacetime:
\begin{equation}
\label{eq:ScalarChargeLag}
\mathcal S_{\rm SC}=\, \frac{1}{2}\int d^D x \left(\, b_1\, F_{Abc}\,F^{Abc} + b_2 \,F_{ab,c}\, F^{ab,c}\right)\,,
\end{equation}
where $b_1 \neq0$, $ b_2 \neq0$. Note that $F_{Aab}$ is symmetric in the transverse indices, while $F_{ab,c}$ is antisymmetric in its first two indices.

The gauge fields can be coupled to matter by introducing two currents $J^A$ and $J^{ab}$ such that, up to total derivatives, the variation of the action has form
\begin{equation}\label{varLAEflatwithJ2}
\delta \mathcal S= \delta \mathcal S_{SC} +\int d^D x\,\left(\delta A_A J^A+ \delta A_{ab} J^{ab}\right) \,.
\end{equation}
Varying the action with respect to $A_{ab}$ and $A_A$ yields the following field equations
\begin{equation}
\begin{aligned}
\delta A_A\,&: \qquad &&b_1 \,\partial_a \partial_b F^{Aab}
=-J^A \,,
\\[5pt]
\delta A_{ab}\,&: &&b_1 \,\partial_A F^{Aab}+ 
b_2\, \partial_c F^{ca,b}=J^{ab} \,.
\end{aligned}
\end{equation}
Combining these equations, one finds a fracton-like continuity equation
\begin{equation}
\partial_A J^A + \partial_a \partial_b J^{ab}=0\,.
\end{equation}
In contrast to the Aristotelian electrodynamics case discussed in subsection \ref{quadracticmodels}, coupling the scalar charge theory to a curved background is non-trivial and requires either imposing constraints on the geometry or the introduction of extra field content \cite{Slagle:2018kqf,Pena-Benitez:2021ipo,Bidussi:2021nmp,Jain:2021ibh,Pena-Benitez:2023aat,Hartong:2024hvs}. Alternatively, we will achieve this coupling below by making use of the real scalar field that compensates for the U(1)/dipole symmetries.

\vspace{8pt}
\noindent\textbf{Fracton field theory:}

\noindent Inspired by the  fracton field theory models of \cite{Pretko:2018jbi,Bidussi:2021nmp}, we consider the following action for a complex scalar field $\Phi$:
\begin{equation}\label{eq:Lfracton}
\mathcal{S}^{HD}_\Phi=\, \frac{1}{2} \int d^D x\, \left( c_1\, \partial_A \Phi \partial^A \Phi^* + c_2\, X_{ab}   X^{ab*} \right)
\end{equation}
where $X_{ab}$ is given by
\begin{equation}
X_{ab} = \partial_a\Phi \partial_b\Phi - \Phi \partial_a \partial_b\Phi\,.
\end{equation}
This action is invariant under the following global symmetry transformations
\begin{equation}
\delta \Phi = \left(i \lambda_{U(1)} + i\lambda_{\rm dp}^a \,x_a + w\lambda_D \right) \Phi\,,
\end{equation}
where $\lambda_{U(1)}$ stands for the $U(1)$ transformation parameter and $\lambda_{\rm dp}^a$ is the parameter of dipole transformations. The global dilatation invariance holds provided $w$ and $z$ are given by \eqref{eq:zandw}.

It is convenient to express the complex scalar field in terms of new variables $\rho$ and $\theta$ corresponding to the radius and the phase of $\Phi$, i.e.
\begin{equation}
\Phi = \rho\,e^{i\theta}\,.
\end{equation}
The action \eqref{eq:Lfracton} can be written as
\begin{equation}
\mathcal{S}^{HD}_\Phi=\ \mathcal{S}^{HD}_\rho +  \mathcal{S}^{HD}_\theta(\rho) \,,
\end{equation}
where $\mathcal{S}^{HD}_\rho$ has been defined in \eqref{eq:LrhoHD} and
\begin{equation}\label{SHDrho}
\mathcal{S}^{HD}_\theta(\rho) =\, \frac{1}{2}\int d^D x\,  \rho^2\left( c_1 \,\partial_A\theta\, \partial^A \theta + c_2\, \rho^2 Z_{ab}Z^{ab}\right)  \,, \quad \text{with} \quad Z_{ab}=\ \partial_a\partial_b \theta \,.
\end{equation}
We will now discuss the gaugings of this model in the three stages mentioned at the beginning of this subsection.

\vspace{8pt}

\noindent (i) \textit{ Gauging the conformal Aristotelian symmetry}: Coupling \eqref{eq:Lfracton} to conformal Aristotelian gravity leads to the curved space action
\begin{equation}\label{eq:LfractonHD}
\mathcal{S}^{cHD}_\Phi=\, \frac{1}{2} \int d^D x\, \Omega\left(c_1\, D_A \Phi D^A \Phi^* + c_2 \, X_{ab} X^{ab*}\right) \,,
\end{equation}
where $X_{ab}$ now reads
\begin{equation}
X_{ab} = D_a\Phi D_b\Phi - \Phi D_{(a} D_{b)}\Phi\,,
\end{equation}
and $D_A \Phi$, $D_a \Phi$, and $D_aD_b\Phi$ are defined like for a real scalar field as in \eqref{eq:CovD} and \eqref{eq:DDphi2}.

In order to gauge-fix the dilatations, we impose the gauge condition \eqref{gaugefixingrho}, which leads to the following action for a real scalar field $\theta$ coupled to Aristotelian gravity
\begin{equation}
\begin{aligned}
\mathcal{S}=\,  \mathcal{S}_{\rm HDAG}+\mathcal{S}_{\theta}^{cHD}  \,,
\end{aligned}
\end{equation}
where $\mathcal{S}_{\rm HDAG}$ is given by \eqref{eq:Agravityhigher} and
\begin{equation}
\mathcal{S}_{\theta}^{cHD} =\ \frac{1}{2} \int d^D x\, \Omega\left(c_1\, \partial_A \theta \partial^A\theta + \, c_2\, Y_{ab}Y^{ab}  \right) \,, \quad \text{with} \quad
Y_{ab} = 
- \mathcal D_{(a} \partial_{b)} \theta
-2 b_{(a} \partial_{b)} \theta
+\delta_{ab}b^c\partial_c \theta 
\,,
\end{equation}
and $\mathcal{D}_a$ is the covariant derivative constructed with the non-conformal spin-connection.

\vspace{8pt}
\noindent (ii) \textit{Gauging the $U(1)$ and dipole symmetries}: Instead of gauging the dilatation symmetry and ending up with a scalar $\theta$ coupled to gravity, we now gauge the  $U(1)$ and dipole symmetries to obtain a higher-derivative theory for a scalar field $\rho$ in flat spacetime. Following \cite{Pretko:2016kxt,Pretko:2018jbi}, in order to gauge the shift symmetry of the angular variable $\theta$ we introduce two independent gauge fields $A_A$ and $A_{ab}$, and consider the action
\begin{align}
\begin{split}
\mathcal{S}=\, \frac{1}{2} \int d^D x\, \Big(& c_1\, \left(\partial_A-iA_A\right) \Phi \left(\partial^A+iA^A\right)  \Phi^* + c_2 \, X_{ab}  
X^{ab*}  \\
&+b_1 \,|\Phi \Phi^*|^{\gamma_1/2}\, F_{Abc}\,F^{Abc} + b_2 \,|\Phi \Phi^*|^{\gamma_2/2}\,F_{ab,c}\, F^{ab,c}\Big) \,,
\end{split}
\end{align}
where $F_{Abc}$ and $F_{ab,c}$ are given in \eqref{eq:ScalarChargeTheoryF} and in this case $X_{ab}$ has the form
\begin{equation}
X_{ab} = \partial_a\Phi \partial_b\Phi - \Phi \partial_a \partial_b\Phi +iA_{ab}\Phi^2\,.
\end{equation}
Under local $U(1)$ transformations and global dilatations, the different fields transform as
\begin{equation}\label{fractontrafos}
\delta \Phi = i\lambda_{U(1)}\Phi +w\lambda_D \Phi, \qquad
\delta A_A= \partial_A \lambda_{U(1)}-z\lambda_D A_A, \qquad
\delta A_{ab}=\partial_a \partial_b \lambda_{U(1)}-2\lambda_D A_{ab} \,,
\end{equation}
where $z$ and $w$ are as given in \eqref{eq:zandw}, and in order for the action to be invariant under dilatation we take
\begin{equation}
\label{eq:Gammas12}
\gamma_1 = 2\frac{3p-9+D}{p-3+D} \,, \qquad \qquad \qquad \gamma_2=2\frac{3p-9-2D}{p-3+D} \,.
\end{equation}
Imposing the gauge-fixing condition \eqref{gaugefixingtheta} we obtain the following action
\begin{align}
\begin{split}
\mathcal{S}=\, \mathcal{S}^{HD}_\rho  + \frac{1}{2} \int d^D x\, \Big(& c_1\, A_A A^A \rho^2 +c_2 \,\rho^4 A_{ab}A^{ab} \\
&+b_1 \,\rho^{\gamma_1}\, F_{Abc}\,F^{Abc} + b_2 \,\rho^{\gamma_2}\,F_{ab,c}\, F^{ab,c}\Big) \,,
\end{split}
\end{align}
that consists of the sum of a higher-derivative action for the real scalar field $\rho$ and a Proca-like version of the scalar charge theory.

\vspace{8pt}
\noindent (iii) \textit{Gauging all symmetries considered in cases (i) and (ii)}: We now wish to couple the gauge fields $A_A\,, A_{ab}$ to conformal Aristotelian gravity. Substituting in the expression for the curvatures the partial derivatives with derivatives that are covariant under all symmetries except $U(1)$, we obtain the following expression for the curvature tensors:
\begin{equation}
 F_{Aab}=D_A A_{ab}-D_{(a} D_{b)} A_A\,, \qquad  F_{ab,c}= D_a A_{bc}- D_b A_{ac}\,.
\end{equation}
Assuming that the gauge fields transform as in eq.~\eqref{fractontrafos}, with $\partial_a \partial_b$ replaced by $D_{(a}\partial_{b)}$, we find that the curvature tensors transform under dilatations and $U(1)$ as follows:
\begin{equation}
\begin{aligned}
\label{eq:ScalarChargeTheoryFTransf}
\delta F_{Aab} =& -(z+2) \lambda_D  F_{Aab}+ D_A D_{(a}\partial_{b)}\lambda_{U(1)} -D_{(a} D_{b)} \partial_A \lambda_{U(1)} \,, \\
\delta F_{ab,c} =& -3 \lambda_D F_{ab,c} + D_a D_{(b}\partial_{c)}\lambda_{U(1)}- D_b D_{(a}\partial_{c)}\lambda_{U(1)} \,.
\end{aligned}
\end{equation}
We see that these curvatures do not transform covariantly under U(1). One way to deal with this situation is to use the compensating scalar $\theta$ to define new curvature tensors that are invariant under $U(1)$. In this way we obtain the following tilded gauge-covariant curvature tensors:
\begin{equation}
\begin{aligned}
\tilde{F}_{Aab}\equiv \, & D_A A_{ab}-D_{(a} D_{b)} A_A  -  D_A D_{(a}\partial_{b)}\theta +D_{(a} D_{b)} \partial_A \theta  \,, \\ 
\tilde{F}_{ab,c}\equiv \, & D_a A_{bc}- D_b A_{ac} -D_a D_{(b}\partial_{c)}\theta + D_b D_{(a}\partial_{c)}\theta \,.
\end{aligned}
\end{equation}
We can then couple the fracton field theory and the scalar charge theory to curved backgrounds by substituting the partial derivatives with covariant derivatives and use the compensating scalar $\rho$ to make the whole action dilatation invariant:
\begin{align}
\label{eq:FractonFieldActioniii}
\begin{split}
\mathcal{S}=\, \frac{1}{2} \int d^D x\, \Omega\Big(&c_1\, \left(\left(D_A-iA_A\right) \Phi \left(D^A+iA^A\right)  \Phi^* + c_2\, X_{ab}  
X^{ab*}\right) \\
&+b_1 \, \rho^{\gamma_1}\, \Tilde{F}_{Abc}\,\Tilde{F}^{Abc} + b_2 \,\rho^{\gamma_2}\,\Tilde{F}_{ab,c}\, \Tilde{F}^{ab,c}\Big) \,.
\end{split}
\end{align}
Here,  $z$ and $w$ are given by eq.~\eqref{zandw-quadcomplexscalar}, the $\gamma$'s are given by by \eqref{eq:Gammas12} and $X_{ab}$ takes the form
\begin{equation}
X_{ab} = D_a\Phi D_b\Phi - \Phi D_{(a} D_{b)}\Phi +iA_{ab}\Phi^2\,,
\end{equation}
where $D_A \Phi$, $D_a \Phi$ and $D_a D_b\Phi$ are defined as in eqs.~\eqref{eq:CovD} and \eqref{eq:DDphi2}. 

Imposing the gauge-fixing condition \eqref{gaugefixPhi} we obtain the total action
\begin{equation}
\mathcal{S}_{\rm Total} = \mathcal{S}_{\rm HDAG} +\mathcal{S}_{\rm Kin} +\mathcal{S}_{\rm Mass} \,,
\end{equation}
where $\mathcal{S}_{\rm HDAG}$ is given in eq.~\eqref{eq:Agravityhigher} and $\mathcal{S}_{\rm Kin}, \mathcal{S}_{\rm Mass}$ are given by
\begin{align} 
\mathcal S_{\rm Kin}=&\, \frac{1}{2}\int d^D x\, \Omega\,\left(b_1 \, F_{Abc}\,F^{Abc} + b_2 \, F_{ab,c}\, F^{ab,c}\right) \,,
\\
\mathcal{S}_{\rm Mass} =&\, \frac{1}{2}\int d^D x \, \Omega\left(c_1\,A_A A^A+c_2\, A_{ab} A^{ab}\right)\,.
\end{align}
We thus end up with another example of a matter coupled Aristotelian gravity theory that consists of the sum of a higher-derivative electric Aristotelian gravity theory and a massive scalar charge theory. A different coupling of the fracton field theory \eqref{eq:Lfracton} to an Aristotelian background and to the scalar charge theory for $p=0$ and $D=3$ has been considered in \cite{Bidussi:2021nmp}.

%%%%%%%%%%%%%%%%%%%%%%%%%%%%%%%%%%%%%
\section{Conclusions}

In this paper we constructed two conformal extensions of the Aristotelian algebra, one embedded as a subalgebra into the other, and showed how they can be used as a first step in constructing matter-coupled Aristotelian gravity theories with no boost symmetry. We distinguished between three different classes of Aristotelian gravity theories: electric, magnetic, and electric-magnetic ones. We use here a nomenclature where `magnetic' refers to the use of an  Aristotelian connection curvature as basic building block in the Lagrangian and `electric' implies that the basic building block in the Lagrangian is an intrinsic torsion tensor. We find a single magnetic Aristotelian gravity Lagrangian but, due to the many intrinsic torsion tensors, there are many electric and electric-magnetic Aristotelian gravity theories. A noteworthy feature of the second-order Aristotelian gravity theories that we considered in this work is that, unlike in Galilei or Carroll gravity, all spin-connections are dependent and hence cannot act as Lagrange multipliers.  Correspondingly, the theory can be constructed without any constraint on the geometry.

We discussed several conformal matter couplings based upon quadratic-derivative and higher-derivative models. The higher-derivative models that we considered were of the type that have been studied in the context of fractons. In these higher-derivative cases we worked with {\sl two} compensating scalars: one for the gravitational dilatations and one for the gauged dipole symmetries (instead of a gravitational dilatation like we did in section 3). The final result we obtained was a matter coupled Aristotelian gravity theory with the matter given by a Lagrangian for the massive dipole gauge fields. 
To obtain more general matter couplings to Aristotelian gravity after gauge-fixing one should replace the compensating scalars by functions of $N$ scalars like in the relativistic case.

It would be interesting to apply our formalism to the massive spin-2 GMP modes \cite{Girvin:1986zz}  in the Fractional Quantum Hall Effect whose helicity has recently been observed \cite{Liang:2024dbb}. A recent theoretical description of these modes has been given but so far without boost symmetry, i.e.~in an Aristotelian context \cite{Gromov:2017qeb}. Using our formalism one could try to construct these massive spin-2 modes in a curved Aristotelian background and to consider the corresponding Schr\"odinger equation in such an Aristotelian background. Other interesting directions to explore are field theories with Aristotelian supersymmetry and the meaning of an Aristotelian string theory. We hope to come back on some of these issues in the nearby future.

%\vskip 2truecm

\section*{Acknowledgments}

We would like to thank Kevin Grosvenor, Giandomenico Palumbo and Matthew Roberts for useful discussions. The work of G.G. has been supported by the predoctoral fellowship FPI-UM R-1006-2021-01. G.G. has also been supported by the COST Action CA22113 for a research stay at the University of Groningen.  The work of E.B. and J.R. is supported by the Croatian Science Foundation project IP-2022-10-5980 “Non-relativistic supergravity and applications'' and by the European Union--NextGenerationEU. P.S.-R. has been supported by a Young Scientist Training Program (YST) fellowship at the Asia Pacific Center for Theoretical Physics (APCTP) through the Science and Technology Promotion Fund and the Lottery Fund of the Korean Government. P.S.-R. has also been supported by the Korean local governments in Gyeongsangbuk-do Province and Pohang City. P.S.-R. acknowledges the Van Swinderen Institute for their hospitality and financial support during the initial stages of this work. The views and opinions expressed are those of the authors only and do not necessarily reflect the official views of the European Union or the European Commission. Neither the European Union nor the European Commission can be held responsible for them.

\vspace{5pt}

\providecommand{\href}[2]{#2}\begingroup\raggedright\endgroup


\begin{thebibliography}{10}

\bibitem{Penrose:1968ar}
R.~Penrose, ``{Structure of space-time},'' in {\em {Battelle Rencontres}},
  pp.~121--235.
\newblock 1968.

\bibitem{Horava:2016vkl}
P.~Horava, ``{Surprises with Nonrelativistic Naturalness},''
  \href{http://dx.doi.org/10.1142/S0218271816450073}{{\em Int. J. Mod. Phys. D}
  {\bfseries 25} no.~13, (2016) 1645007},
  \href{http://arxiv.org/abs/1608.06287}{{\ttfamily arXiv:1608.06287
  [hep-th]}}.

\bibitem{Yan:2017mse}
Z.~Yan, {\em {Nonrelativistic Naturalness in Aristotelian Quantum Field
  Theories}}.
\newblock PhD thesis, UC, Berkeley (main), 2017.

\bibitem{Novak:2019wqg}
I.~Novak, J.~Sonner, and B.~Withers, ``{Hydrodynamics without boosts},''
  \href{http://dx.doi.org/10.1007/JHEP07(2020)165}{{\em JHEP} {\bfseries 07}
  (2020) 165}, \href{http://arxiv.org/abs/1911.02578}{{\ttfamily
  arXiv:1911.02578 [hep-th]}}.

\bibitem{deBoer:2020xlc}
J.~de~Boer, J.~Hartong, E.~Have, N.~A. Obers, and W.~Sybesma, ``{Non-Boost
  Invariant Fluid Dynamics},''
  \href{http://dx.doi.org/10.21468/SciPostPhys.9.2.018}{{\em SciPost Phys.}
  {\bfseries 9} no.~2, (2020) 018},
  \href{http://arxiv.org/abs/2004.10759}{{\ttfamily arXiv:2004.10759
  [hep-th]}}.

\bibitem{Armas:2020mpr}
J.~Armas and A.~Jain, ``{Effective field theory for hydrodynamics without
  boosts},'' \href{http://dx.doi.org/10.21468/SciPostPhys.11.3.054}{{\em
  SciPost Phys.} {\bfseries 11} no.~3, (2021) 054},
  \href{http://arxiv.org/abs/2010.15782}{{\ttfamily arXiv:2010.15782
  [hep-th]}}.

\bibitem{Marotta:2023ayw}
V.~E. Marotta and R.~J. Szabo, ``{Godbillon-Vey invariants of Non-Lorentzian
  spacetimes and Aristotelian hydrodynamics},''
  \href{http://dx.doi.org/10.1088/1751-8121/acfc07}{{\em J. Phys. A} {\bfseries
  56} no.~45, (2023) 455201}, \href{http://arxiv.org/abs/2304.12722}{{\ttfamily
  arXiv:2304.12722 [hep-th]}}.

\bibitem{Grosvenor:2024vcn}
K.~T. Grosvenor, N.~A. Obers, and S.~P. Patil, ``{Hydrodynamics without
  Boost-Invariance from Kinetic Theory: From Perfect Fluids to Active
  Flocks},'' \href{http://arxiv.org/abs/2501.00025}{{\ttfamily arXiv:2501.00025
  [hep-th]}}.

\bibitem{Gubser:2010ze}
S.~S. Gubser, ``{Symmetry constraints on generalizations of Bjorken flow},''
  \href{http://dx.doi.org/10.1103/PhysRevD.82.085027}{{\em Phys. Rev. D}
  {\bfseries 82} (2010) 085027},
  \href{http://arxiv.org/abs/1006.0006}{{\ttfamily arXiv:1006.0006 [hep-th]}}.

\bibitem{Chattopadhyay:2018apf}
C.~Chattopadhyay, U.~Heinz, S.~Pal, and G.~Vujanovic, ``{Higher order and
  anisotropic hydrodynamics for Bjorken and Gubser flows},''
  \href{http://dx.doi.org/10.1103/PhysRevC.97.064909}{{\em Phys. Rev. C}
  {\bfseries 97} no.~6, (2018) 064909},
  \href{http://arxiv.org/abs/1801.07755}{{\ttfamily arXiv:1801.07755
  [nucl-th]}}.

\bibitem{Billo:2016cpy}
M.~Bill{\`o}, V.~Gon{\c{c}}alves, E.~Lauria, and M.~Meineri, ``{Defects in
  conformal field theory},''
  \href{http://dx.doi.org/10.1007/JHEP04(2016)091}{{\em JHEP} {\bfseries 04}
  (2016) 091}, \href{http://arxiv.org/abs/1601.02883}{{\ttfamily
  arXiv:1601.02883 [hep-th]}}.

\bibitem{Arav:2024exg}
I.~Arav, J.~P. Gauntlett, Y.~Jiao, M.~M. Roberts, and C.~Rosen,
  ``{Superconformal monodromy defects in $ \mathcal{N} $=4 SYM and LS
  theory},'' \href{http://dx.doi.org/10.1007/JHEP08(2024)177}{{\em JHEP}
  {\bfseries 08} (2024) 177}, \href{http://arxiv.org/abs/2405.06014}{{\ttfamily
  arXiv:2405.06014 [hep-th]}}.

\bibitem{Gromov:2017qeb}
A.~Gromov and D.~T. Son, ``{Bimetric Theory of Fractional Quantum Hall
  States},'' \href{http://dx.doi.org/10.1103/PhysRevX.7.041032}{{\em Phys. Rev.
  X} {\bfseries 7} no.~4, (2017) 041032},
  \href{http://arxiv.org/abs/1705.06739}{{\ttfamily arXiv:1705.06739
  [cond-mat.str-el]}}. [Addendum: Phys.Rev.X 8, 019901 (2018)].

\bibitem{Chamon:2004lew}
C.~Chamon, ``{Quantum Glassiness},''
  \href{http://dx.doi.org/10.1103/physrevlett.94.040402}{{\em Phys. Rev. Lett.}
  {\bfseries 94} no.~4, (2005) 040402},
  \href{http://arxiv.org/abs/cond-mat/0404182}{{\ttfamily
  arXiv:cond-mat/0404182}}.

\bibitem{Haah:2011drr}
J.~Haah, ``{Local stabilizer codes in three dimensions without string logical
  operators},'' \href{http://dx.doi.org/10.1103/physreva.83.042330}{{\em Phys.
  Rev. A} {\bfseries 83} no.~4, (2011) 042330},
  \href{http://arxiv.org/abs/1101.1962}{{\ttfamily arXiv:1101.1962
  [quant-ph]}}.

\bibitem{Pretko:2016kxt}
M.~Pretko, ``{Subdimensional Particle Structure of Higher Rank U(1) Spin
  Liquids},'' \href{http://dx.doi.org/10.1103/PhysRevB.95.115139}{{\em Phys.
  Rev. B} {\bfseries 95} no.~11, (2017) 115139},
  \href{http://arxiv.org/abs/1604.05329}{{\ttfamily arXiv:1604.05329
  [cond-mat.str-el]}}.

\bibitem{Pretko:2018jbi}
M.~Pretko, ``{The Fracton Gauge Principle},''
  \href{http://dx.doi.org/10.1103/PhysRevB.98.115134}{{\em Phys. Rev. B}
  {\bfseries 98} no.~11, (2018) 115134},
  \href{http://arxiv.org/abs/1807.11479}{{\ttfamily arXiv:1807.11479
  [cond-mat.str-el]}}.

\bibitem{Slagle:2018kqf}
K.~Slagle, A.~Prem, and M.~Pretko, ``{Symmetric Tensor Gauge Theories on Curved
  Spaces},'' \href{http://dx.doi.org/10.1016/j.aop.2019.167910}{{\em Annals
  Phys.} {\bfseries 410} (2019) 167910},
  \href{http://arxiv.org/abs/1807.00827}{{\ttfamily arXiv:1807.00827
  [cond-mat.str-el]}}.

\bibitem{Pena-Benitez:2021ipo}
F.~Pe\~na Benitez, ``{Fractons, symmetric gauge fields and geometry},''
  \href{http://dx.doi.org/10.1103/PhysRevResearch.5.013101}{{\em Phys. Rev.
  Res.} {\bfseries 5} no.~1, (2023) 013101},
  \href{http://arxiv.org/abs/2107.13884}{{\ttfamily arXiv:2107.13884
  [cond-mat.str-el]}}.

\bibitem{Bidussi:2021nmp}
L.~Bidussi, J.~Hartong, E.~Have, J.~Musaeus, and S.~Prohazka, ``{Fractons,
  dipole symmetries and curved spacetime},''
  \href{http://dx.doi.org/10.21468/SciPostPhys.12.6.205}{{\em SciPost Phys.}
  {\bfseries 12} no.~6, (2022) 205},
  \href{http://arxiv.org/abs/2111.03668}{{\ttfamily arXiv:2111.03668
  [hep-th]}}.

\bibitem{Jain:2021ibh}
A.~Jain and K.~Jensen, ``{Fractons in curved space},''
  \href{http://dx.doi.org/10.21468/SciPostPhys.12.4.142}{{\em SciPost Phys.}
  {\bfseries 12} no.~4, (2022) 142},
  \href{http://arxiv.org/abs/2111.03973}{{\ttfamily arXiv:2111.03973
  [hep-th]}}.

\bibitem{Pena-Benitez:2023aat}
F.~Pe\~na Ben\'\i{}tez and P.~Salgado-Rebolledo, ``{Fracton gauge fields from
  higher-dimensional gravity},''
  \href{http://dx.doi.org/10.1007/JHEP04(2024)009}{{\em JHEP} {\bfseries 04}
  (2024) 009}, \href{http://arxiv.org/abs/2310.12610}{{\ttfamily
  arXiv:2310.12610 [hep-th]}}.

\bibitem{Hartong:2024hvs}
J.~Hartong, G.~Palumbo, S.~Pekar, A.~P\'erez, and S.~Prohazka, ``{Fractons on
  curved spacetime in 2 + 1 dimensions},''
  \href{http://dx.doi.org/10.21468/SciPostPhys.18.1.022}{{\em SciPost Phys.}
  {\bfseries 18} no.~1, (2025) 022},
  \href{http://arxiv.org/abs/2409.04525}{{\ttfamily arXiv:2409.04525
  [hep-th]}}.

\bibitem{Figueroa-OFarrill:2018ilb}
J.~Figueroa-O'Farrill and S.~Prohazka, ``{Spatially isotropic homogeneous
  spacetimes},'' \href{http://dx.doi.org/10.1007/JHEP01(2019)229}{{\em JHEP}
  {\bfseries 01} (2019) 229}, \href{http://arxiv.org/abs/1809.01224}{{\ttfamily
  arXiv:1809.01224 [hep-th]}}.

\bibitem{Figueroa-OFarrill:2020gpr}
J.~Figueroa-O'Farrill, ``{On the intrinsic torsion of spacetime structures},''
  \href{http://arxiv.org/abs/2009.01948}{{\ttfamily arXiv:2009.01948
  [hep-th]}}.

\bibitem{Ghosh:2025eob}
J.~K. Ghosh, F.~Pe\~na Ben\'\i{}tez, and P.~Salgado-Rebolledo, ``{Hydrostatic
  equilibrium in multi-Weyl semimetals},''
  \href{http://arxiv.org/abs/2504.20361}{{\ttfamily arXiv:2504.20361
  [cond-mat.str-el]}}.

\bibitem{Bergshoeff:2023rkk}
E.~Bergshoeff, J.~Figueroa-O'Farrill, K.~van Helden, J.~Rosseel, I.~Rotko, and
  T.~ter Veldhuis, ``{$p$-brane Galilean and Carrollian geometries and
  gravities},'' \href{http://dx.doi.org/10.1088/1751-8121/ad4c62}{{\em J. Phys.
  A} {\bfseries 57} no.~24, (2024) 245205},
  \href{http://arxiv.org/abs/2308.12852}{{\ttfamily arXiv:2308.12852
  [hep-th]}}.

\bibitem{Freedman:2012zz}
D.~Z. Freedman and A.~Van~Proeyen,
  \href{http://dx.doi.org/10.1017/CBO9781139026833}{{\em {Supergravity}}}.
\newblock Cambridge Univ. Press, Cambridge, UK, 5, 2012.

\bibitem{Bergshoeff:2024ilz}
E.~A. Bergshoeff, P.~Concha, O.~Fierro, E.~Rodr{\'\i}guez, and J.~Rosseel, ``{A
  conformal approach to Carroll gravity},''
  \href{http://dx.doi.org/10.1007/JHEP07(2025)075}{{\em JHEP} {\bfseries 07}
  (2025) 075}, \href{http://arxiv.org/abs/2412.17752}{{\ttfamily
  arXiv:2412.17752 [hep-th]}}.

\bibitem{Henneaux:2021yzg}
M.~Henneaux and P.~Salgado-Rebolledo, ``{Carroll contractions of
  Lorentz-invariant theories},''
  \href{http://dx.doi.org/10.1007/JHEP11(2021)180}{{\em JHEP} {\bfseries 11}
  (2021) 180}, \href{http://arxiv.org/abs/2109.06708}{{\ttfamily
  arXiv:2109.06708 [hep-th]}}.

\bibitem{Hansen:2021fxi}
D.~Hansen, N.~A. Obers, G.~Oling, and B.~T. S\o{}gaard, ``{Carroll Expansion of
  General Relativity},''
  \href{http://dx.doi.org/10.21468/SciPostPhys.13.3.055}{{\em SciPost Phys.}
  {\bfseries 13} no.~3, (2022) 055},
  \href{http://arxiv.org/abs/2112.12684}{{\ttfamily arXiv:2112.12684
  [hep-th]}}.

\bibitem{Figueroa-OFarrill:2022mcy}
J.~Figueroa-O'Farrill, E.~Have, S.~Prohazka, and J.~Salzer, ``{The gauging
  procedure and carrollian gravity},''
  \href{http://dx.doi.org/10.1007/JHEP09(2022)243}{{\em JHEP} {\bfseries 09}
  (2022) 243}, \href{http://arxiv.org/abs/2206.14178}{{\ttfamily
  arXiv:2206.14178 [hep-th]}}.

\bibitem{Bergshoeff:2022eog}
E.~Bergshoeff, J.~Figueroa-O'Farrill, and J.~Gomis, ``{A non-lorentzian
  primer},'' \href{http://dx.doi.org/10.21468/SciPostPhysLectNotes.69}{{\em
  SciPost Phys. Lect. Notes} {\bfseries 69} (2023) 1},
  \href{http://arxiv.org/abs/2206.12177}{{\ttfamily arXiv:2206.12177
  [hep-th]}}.

\bibitem{Girvin:1986zz}
S.~M. Girvin, A.~H. MacDonald, and P.~M. Platzman, ``{Magneto-roton theory of
  collective excitations in the fractional quantum Hall effect},''
  \href{http://dx.doi.org/10.1103/PhysRevB.33.2481}{{\em Phys. Rev. B}
  {\bfseries 33} (1986) 2481--2494}.

\bibitem{Liang:2024dbb}
J.~Liang, Z.~Liu, Z.~Yang, Y.~Huang, U.~Wurstbauer, C.~R. Dean, K.~W. West,
  L.~N. Pfeiffer, L.~Du, and A.~Pinczuk, ``{Evidence for chiral graviton modes
  in fractional quantum Hall liquids},''
  \href{http://dx.doi.org/10.1038/s41586-024-07201-w}{{\em Nature} {\bfseries
  628} no.~8006, (2024) 78--83}.

\end{thebibliography}
\end{document}